\documentclass[10pt,a4paper,twocolumn]{IEEEtran}
\usepackage{diagbox}
\usepackage{amssymb}
\usepackage[english]{babel}
\usepackage{graphicx}
\usepackage{latexsym}
\usepackage{varioref}
\usepackage{amssymb}
\usepackage[english]{babel}
\usepackage{graphicx}
\usepackage{latexsym}
\usepackage{varioref}
\usepackage{amsmath}
\usepackage{mathrsfs}
\usepackage{array}
\usepackage{hhline}
\usepackage{dcolumn}
\usepackage{tabularx}
\usepackage{algorithm}
\usepackage{algorithmic}
\usepackage{calc}
\usepackage{subfigure}
\usepackage{amssymb,amsmath,epsfig}
\usepackage[usenames]{color}
\usepackage{amssymb,array,multirow,bigdelim}
\usepackage{t1enc}
\usepackage{pstricks,pst-node,psfrag,pst-plot}
\usepackage{graphicx,parskip,amsbsy}
\usepackage{boxedminipage}
\usepackage{indentfirst}
\usepackage{booktabs}
\usepackage[noadjust]{cite}
\usepackage{graphicx}
\usepackage{latexsym}
\usepackage{varioref}
\usepackage{epsfig}
\usepackage{caption}
\usepackage{amsmath,amssymb}
\usepackage{array}
\usepackage{hhline}
\usepackage{dcolumn}
\usepackage{multirow}
\usepackage{tabularx}
\usepackage{algorithm}
\usepackage{algorithmic}
\usepackage{calc}
\usepackage{subfigure}
\usepackage{amssymb,amsmath,epsfig}
\usepackage[usenames]{color}
\usepackage{t1enc}
\usepackage{pstricks,pst-node,psfrag,pst-plot}
\usepackage{graphicx,parskip,amsbsy}
\usepackage{boxedminipage}
\usepackage{indentfirst}
\usepackage{booktabs}
\usepackage{clrscode}
\renewcommand{\blue}{\black}
\renewcommand{\red}{\black}
\newcommand{\tabincell}[2]{\begin{tabular}{@{}#1@{}}#2\end{tabular}}% change line in table cell
 \graphicspath{
{./new_figures/}
}

\newcommand{\bs}{\boldsymbol}

\newcommand{\bsTheta}{{\bs \Theta}}

\newcommand{\rem}[1]{}
\newcommand{\addnew}[1]{{\color{red} #1}}
\providecommand{\texlivekeywords}[1]{\textbf{\textit{Index terms---}} #1}
\begin{document}
	
	\pagestyle{plain}
	\title{An Empirical Air-to-Ground Channel Model Based\\on Passive Measurements in LTE}
    		\author{ Xuesong Cai,~\IEEEmembership{Member,~IEEE,} Jos\'e Rodr\'iguez-Pi\~neiro,~\IEEEmembership{Member,~IEEE,} Xuefeng Yin,~\IEEEmembership{Member,~IEEE,} \\Bo Ai,~\IEEEmembership{Senior Member,~IEEE,} Gert Fr{\o}lund Pedersen,~\IEEEmembership{Senior Member,~IEEE,} and Antonio P\'erez Yuste,~\IEEEmembership{Senior Member,~IEEE}
		
		\thanks{
}
	}
	
	\maketitle \thispagestyle{plain}
	
\begin{abstract}
In this paper, a recently conducted measurement campaign for unmanned-aerial-vehicle (UAV) channels is introduced. The downlink signals of an in-service long-time-evolution (LTE) network which is deployed in a suburban scenario were acquired. Five horizontal and five vertical flight routes were considered. The channel impulse responses (CIRs) are extracted from the received data by exploiting the cell specific signals (CRSs). Based on the CIRs, the parameters of multipath components (MPCs) are estimated by using a high-resolution algorithm derived according to the space-alternating generalized expectation-maximization (SAGE) principle. Based on the SAGE results, channel characteristics including the path loss, shadow fading, fast fading, delay spread and Doppler frequency spread are thoroughly investigated for different heights and horizontal distances, which constitute a stochastic model.

\end{abstract}
\texlivekeywords{
Unmanned aerial vehicle, air-to-ground channel, path loss, delay spread, and Doppler frequency spread.
}

\IEEEpeerreviewmaketitle

\section{Introduction}
Historically, unmanned aerial vehicles (UAVs) were mainly used for military operations in hostile environments for safety reasons \cite{zeng2016wireless,valavanis2014handbook}. Due to the decrease in the cost and size, UAVs are being more accessible for general-purpose civil and commercial applications, such as video surveillance, weather monitoring, search and rescue operations, precision farming, wildlife monitoring, and transportation, among others \cite{us2013unmanned}. The network service restoration after infrastructure damage in natural disasters or base station (BS) relaying in crowded areas (one of the key scenarios addressed by fifth generation (5G) communication systems \cite{zeng2016wireless,osseiran2014scenarios}) also fit well \cite{zhan2011wireless,zhou2015multi}.

Different communication links can be involved in the aforementioned applications of UAVs, such as:
\begin{itemize}
	\item Air-to-ground (A2G) communication channel between a ground BS and the UAV. This communication channel can serve to many different purposes, such as to give support to control and non-payload communications (CNPC) or communication and control link (C2) (see \cite{european2013roadmap,zte2017consideration,alcatel2017evaluation,us2013integration}), to serve data traffic for UAV-based applications (e.g., transmitting video surveillance data from UAVs to ground) or to relay traffic of a BS. Hence, the requirements in terms of availability, quality of service, latency or throughput can be very different. Moreover, the nature of the propagation channel is very different from that of historically considered A2G links, e.g., the ones for civil aviation, with elevated ground site antennas in open areas and narrowband signals with very high transmit signal power and short duty cycles with no continuous reception required \cite{matolak2012air}.

%Moreover, nature of UAV communications is very different from the one of historically considered A2G links like the ones for civil aviation (elevated ground site antennas in open areas, narrowband signals with very high transmit signal power and short duty cycles with no continuous reception required \cite{matolak2012air})
	\item A2G communication channel between a mobile terminal at ground and the UAV. This communication channel would serve mainly for station relaying in crowded areas or service restoration after infrastructure damage in natural disasters.
	\item Air-to-air (A2A) communication channel between flying UAVs. Communications between flying UAVs make sense under the scenarios of relaying communications, cooperative air control strategies between UAVs, collision-avoidance systems, etc.
\end{itemize}

It is noteworthy that %different from being independent communication links,
many potential future applications for UAVs would simultaneously rely on several of the aforementioned communication channels. As an example, in \cite{zhou2015multi}, an aerial-ground cooperative vehicular networking architecture is proposed, considering both A2A and A2G communications.

\subsection{Motivation}

In many countries, the UAVs are limited in visual line-of-sight (LoS) with maximum heights at about $100$\,m and $150$\,m \cite{amorim2017radio,european2013roadmap}. Under these constraints, cellular networks pose themselves as natural candidates to give support to CNPC or C2 communications as well as to serve data traffic for UAV-based applications, which has attracted significant research attentions.  It makes sense to ask whether an off-the-shelf commercial cellular network could be applied in this scenario. The 3rd Generation Partnership Project (3GPP) has approved corresponding study item \cite{LTEUAVSI} and working item \cite{LTEUAVWI} to investigate the feasibility of serving aerial vehicles using long-time-evolution (LTE) network deployments with BS antennas targeting terrestrial coverage.

The A2G communication channel between a ground BS and the UAV is very different from the ones conventionally considered for terrestrial communication systems. Intuitively, the channel exhibits more clearance (more LoS-like) compared to that for ground users \cite{welch2016evolving} (which could give special attention to millimeter wave (mmWave) bands for future high throughput UAV-based applications \cite{khawaja2017uav,rupasinghe2016optimum}). However, the channel clearance implies a higher level of interference from neighboring BSs \cite{van2016lte}, and the uplink communications from the UAV also affect neighboring BSs, distorting ground mobile users \cite{lin2017sky} and other UAVs nearby in the UAV-swarming cases. {\blue Moreover, the BS antennas of commercial cellular networks are usually with directional beams and tilted downwards targeting terrestrial coverage, which has significant influence on the A2G channel characterization in reality.} A comprehensive understanding of the A2G channel is essential for facilitating the design and performance evaluation of wireless techniques for UAV applications, e.g., power control \cite{ZTE3gpppowercontrol}, interference cancellation \cite{ZTE3gppinterference}, etc. Different channel modeling approaches are required, e.g., considering the influence of higher user-equipment heights that are not typical in terrestrial communications. Therefore, in this contribution by exploiting a commercial LTE system in operation, the characteristics of the A2G channel {\red in a realistic LTE network} are thoroughly investigated in a suburban environment in Shanghai, China.

\subsection{Related Work}

In the following paragraphs related papers are considered, indicating which studies are based on simulations or measurements. The main aspects of the considered research are specified.

\textbf{A2G channel characterization by simulations}

{\red In \cite{lin2017sky}, the authors studied the LTE connectivity for small and low altitude UAVs at different heights through simulations. In \cite{feng2006path} new path loss models for A2G channels in an urban environment were proposed for frequencies between $200$\,MHz and $5$\,GHz. The investigations in \cite{tameh19973,khawaja2017uav} exploited the ray-tracing tools for characterizing the large-scale channel parameters. The geometry-based stochastic channel modeling approach was also applied in \cite{schneckenburger2017modeling,wentz2015mimo}.}

 %The influence of the elevation angle on the path loss and shadowing was evaluated by a three-dimensional outdoor deterministic ray-tracing model \cite{tameh19973}. {\red The geometry-based stochastic channel modeling approach was exploited in \cite{schneckenburger2017modeling,wentz2015mimo}. In \cite{khawaja2017uav}, mmWave bands ($28$\,GHz and $60$\,GHz) were considered based on ray-tracing simulations.} Although no measurement results were included in this case, a Universal Software-defined Radio Peripheral (USRP) based channel sounder suitable to be used for A2G characterizations at $60$\,GHz was presented.}

%In \cite{schneckenburger2017modeling}, a geometry-based stochastic channel modeling approach for the A2G channel in L-band was considered, whereas the authors showed how the channel parameters can be derived from measurement data. Geometry-based stochastic channel model was also applied in \cite{wentz2015mimo} to investigate the multiple-input-multiple-output (MIMO) characteristics. In \cite{khawaja2017uav}, mmWave bands ($28$\,GHz and $60$\,GHz) were considered, based on ray-tracing simulations for urban, suburban, rural and over sea environments. Although no measurement results were included in this case, a Universal Software-defined Radio Peripheral (USRP) based channel sounder suitable to be used for A2G characterizations at $60$\,GHz was presented.

\textbf{A2G channel characterization by measurements}

Low altitude (below 300\,m) A2G channels have been investigated based on measurements in \cite{amorim2017radio,al2017modeling,khawaja2016uwb,yanmaz2011channel}. The researches in \cite{amorim2017radio,al2017modeling} exploited the cellular networks at the frequency bands at around 800 MHz. Narrowband characteristics, i.e., path loss and shadow fading were considered. It is noteworthy that the authors in \cite{al2017modeling} characterized the path loss for the UAV links as an excess value to the path loss that would correspond to a terrestrial user. The proposed path loss expression depends on the angle of the UAV with respect to the BS, which is a composite effect of the propagation channel and the empirical configuration of BS antenna. In \cite{khawaja2016uwb}, statistical models characterizing large-scale fading, small-scale fading and multipath propagation were proposed at the frequency range between $3.1$\,GHz and $5.3$\,GHz. The blockage of tree foliage was also investigated. In \cite{yanmaz2011channel}, modeling of path loss exponents for A2G links in open field and campus scenarios was conducted. The effects of the UAV orientations were also observed. {\red Regarding the high altitude (above 300\,m) A2G channel characterization, the richest set of recent measurements can be found in \cite{matolak2015air,matolak2014air,matolak2016air,sun2017air,matolak2017air,matolak2017air2,matolak2014initial,matolak2013air,matolak2015unmanned,matolak2012air,matolak2014air2}, where the L and C bands were considered. Channel parameters such as path loss, delay spread, stationarity distance, K-factor and interband and spatial correlation were evaluated for different environments. Different from the low altitude investigations, the aircraft used in these measurements was large in size and flied up to 50\,km of altitude with high speeds from $70$\,m/s to $112$\,m/s. In addition, the interest in characterizing multiple-input-multiple-output (MIMO) A2G channels was also appreciated in \cite{willink2016measurement,kung2010measuring,newhall2003wideband}.}

{\scriptsize
\begin{table*}%\begin{sidewaystable}
\centering
\begin{small}
\begin{tabular} {cccccccc}
\hline\hline %\toprule
\scriptsize Measurements & \scriptsize \tabincell{c}{Carrier\\frequency}  & \scriptsize \tabincell{c}{UAV\\type}   & \scriptsize Altitude [m]    & \scriptsize Antenna & \scriptsize \tabincell{c}{Frequency\\selectivity}  & \scriptsize Environment &  \scriptsize \tabincell{c}{Channel\\characterisitics}
\\ \hline %%\midrule
\scriptsize Ref. \cite{amorim2017radio}  & \scriptsize 800 MHz   & \scriptsize small UAV  & \scriptsize $<$ 120  & \scriptsize SISO & \scriptsize narrowband  & \scriptsize LTE network  & \scriptsize path loss, shadowing \\
\scriptsize Ref. \cite{al2017modeling}  &  \scriptsize 850 MHz  &  \scriptsize small UAV   &    \scriptsize $<$ 120    &   \scriptsize SISO  & \scriptsize narrowband  & \scriptsize suburban  & \scriptsize path loss    \\
\scriptsize Ref. \cite{khawaja2016uwb}    & \scriptsize 3.1 GHz - 5.3 GHz    & \scriptsize small UAV   & \scriptsize  $<$ 16 & \scriptsize SISO  & \scriptsize wideband  & \scriptsize open, suburban & \scriptsize \tabincell{c}{large-scale fading, small-scale fading,\\ multipath, foliage blockage} \\
   \scriptsize Ref. \cite{yanmaz2011channel}  & \scriptsize  2.4 GHz             &  \scriptsize small UAV   &   \scriptsize $<$ 120     &  \scriptsize SISO            &\scriptsize narrowband  & \scriptsize open, campus   & \scriptsize \tabincell{c}{path loss, antenna orientation}   \\ %\midrule
\scriptsize Refs. \cite{matolak2015air,matolak2014air,matolak2016air,sun2017air,matolak2017air,matolak2017air2,matolak2014initial,matolak2013air,matolak2015unmanned,matolak2012air}  &   \scriptsize 965 MHz, 5 GHz  &  \scriptsize aircraft  &  \scriptsize 500 - 2000  &  \scriptsize SISO  & \scriptsize wideband & \scriptsize \tabincell{c}{over-water, hilly,\\ mountain, suburban}& \scriptsize \tabincell{c}{path loss, delay spread,\\ stationarity, K-factor, spatial correlation}\\
  \scriptsize Ref. \cite{schneckenburger2016measurement}   & \scriptsize 970 MHz   &   \scriptsize aircraft   &   \scriptsize 3000 - 9000   &  \scriptsize SISO          & \scriptsize wideband   & \scriptsize \tabincell{c}{en-route cruise,\\ climb-and-descending,\\ takeoff-and-landing}& \scriptsize \tabincell{c}{power delay profile,\\ Doppler delay profile}\\
  \scriptsize Ref. \cite{willink2016measurement}    & \scriptsize 915 MHz      &     \scriptsize small UAV    & \scriptsize 200   &    \scriptsize MIMO               & \scriptsize wideband & \scriptsize suburban & \scriptsize \tabincell{c}{delay spread,\\ spatial correlation}  \\ %\midrule

\scriptsize Ref. \cite{kung2010measuring}    & \scriptsize 2.4 GHz    & \scriptsize small UAV   & \scriptsize \scriptsize 75     & \scriptsize MIMO       & \scriptsize narrowband  & \scriptsize - & \scriptsize \tabincell{c}{diversity,\\ antenna correlation} \\
\hline%\botrule
\end{tabular}
%\vspace{0.6cm}\\
%{\scriptsize SS：郊区街道； EW：高速公路； UC: 城市峡谷； Micro: 宏小区； Pico: 微微小区； H(L)VTD：高（低）{车流量}； }\\
%{\scriptsize PDP：功率时延谱； DD：多普勒-时延； PSD：
%功率谱密度； STF：空间-时间-频率 CF：相关函数； LCR：水平交叉率； SDF：空间-多普勒-频率}\\
%{\scriptsize PDF：概率密度函数； PL：{路径衰落}； CDF：累积分布函数； CT相关时间。}
%\rput[tr]{0}(19.8,-9.1){\begin{minipage}[b]{16cm}
%{\scriptsize SS：郊区街道； EW：高速公路； UC: 城市峡谷； Micro: 宏小区； Pico: 微微小区； H(L)VTD：高（低）{车流量}； }\\
%{\scriptsize PDP：功率时延谱； DD：多普勒-时延； PSD：
%功率谱密度； STF：空间-时间-频率 CF：相关函数； LCR：水平交叉率； SDF：空间-多普勒-频率}\\
%{\scriptsize PDF：概率密度函数； PL：{路径衰落}； CDF：累积分布函数； CT相关时间。}  \end{minipage}}
\end{small}
\caption{Important A2G UAV channel measurement campaigns.\label{TableQ}}
\end{table*}
}

Table \ref{TableQ} summarizes important measurement campaigns for UAV A2G channels. It can be concluded that although there is an increasing interest on the use of UAVs for civil and commercial applications, not many researches focus on the characterization for the A2G channel in scenarios similar to those being expected in most practical applications and UAV regulations, i.e., with small UAVs and low heights. Most of the research activities are related to high altitude A2G channels with large UAVs (e.g., aircrafts) and high speeds. For low altitude A2G channel investigations, it is deficient in characterizing the wideband characteristics. Comprehensive investigation for the low altitude A2G channel between individual BSs of an, e.g., LTE network and an UAV with typical size is still in necessity.

\subsection{Main Contribution and Novelties}

Motivated by the above background and research gaps, a comprehensive modeling work for the low altitude A2G channel is conducted in this paper. The main contributions and novelties are:

%\note{here we have some aspects (not written yet) as the main differences between other works and ours (each of the items need to be more ellaborated after including all the papers in the state of the art part)}
%\begin{itemize}
%%	\item \note{use of lte commercial network}
%%	\item \note{study of the influence of the distance and the height in a sistematic way}
%	\item \note{study of speed, distance and height values according to the expected uses of UAVs and the current regulations in most of the countries}
%	\item \note{use of a typical-sized UAV (not an aircraft, e.g. allowing for arbitrary flight dynamics) provided with a good quality receiver eligible for research purposes (not a commercial phone)}
%	\item \note{things that you can say regarding advantages of using sage}
%\end{itemize}

\begin{itemize}
	\item A passive channel sounding approach was applied. The downlink signals of a commercial LTE network were collected and used for extracting the channel impulse responses (CIRs). The measurements can be conducted conveniently by using the passive sounding approach. The channel characteristics own rigid fidelity to those experienced by the user equipments on board UAV, since the configurations including the carrier frequency, bandwidth, antenna, UAV type, etc. are exactly the ones applied in a commercial LTE network. Moreover, the 18\,MHz bandwidth of the LTE downlink signals allows analyzing not only the narrowband characteristics, but also the wideband behaviours of the A2G channels, e.g., the joint channel dispersions in delay and Doppler frequency domains.

	%\item The measurement campaign included five horizontal flights at different heights and five vertical flights with different horizontal distances, in a suburban scenario in Shanghai, China. A high resolution parameter estimation algorithm is derived based on the space-alternating generalized expectation-maximization (SAGE) principle to estimate the delays, Doppler frequencies, and complex attenuations of multipath components (MPC). The channel behaviours and corresponding physical mechanisms were investigated carefully at MPC level.

\item {\red The measurement campaign included five horizontal flights at different heights and five vertical flights with different horizontal distances, in a suburban scenario in Shanghai, China. Delays, Doppler frequencies, and complex attenuations of multipath components (MPC) were estimated. The channel behaviours and corresponding physical mechanisms were investigated carefully at MPC level.} Characteristics including path loss, shadow fading, fast fading, delay spread and Doppler frequency spread are comprehensively investigated for different flights. The dependencies of the channel characteristics on the height and the horizontal distance are studied in a systematical way. The resulted channel model would be useful to investigate the inter-A2G channel interference among BSs.
%\item {\red The measurement campaign included five horizontal flights at different heights and five vertical flights with different horizontal distances, in a suburban scenario in Shanghai, China. A high resolution parameter estimation algorithm is derived based on the space-alternating generalized expectation-maximization (SAGE) principle to estimate the delays, Doppler frequencies, and complex attenuations of multipath components (MPC). The channel behaviours and corresponding physical mechanisms were investigated carefully at MPC level. }Based on the results of the SAGE algorithm, channel characteristics including path loss, shadow fading, fast fading, delay spread and Doppler frequency spread are comprehensively investigated for different flights. The dependencies of the channel characteristics on the height and the horizontal distance are studied in a systematical way. The resulted channel model would be useful to investigate the inter-A2G channel interference among BSs.
\end{itemize}

\section{Measurements and raw data processing}\label{sect:campaigns}
In this section, the measurement equipment, scenario and the experiment specifications, including the flight routes, carrier frequency and bandwidth of sounding signals, receiver antenna etc. are described. Furthermore, the procedures for extracting CIRs from the received raw data and then estimating the parameters of MPCs based on the CIRs are elaborated.

\subsection{Measurement equipment}\label{section:equipment}
\begin{figure}
\centering
\psfrag{A}[c][c][0.6]{GPS disciplined oscillator}
\psfrag{B}[r][r][0.4]{Ref in}
\psfrag{C}[c][c][0.6]{USRP}
\psfrag{D}[c][c][0.6]{Small computer}
\psfrag{E}[c][c][0.6]{Router}
\psfrag{F}[l][l][0.6]{Discone antenna}
\psfrag{G}[l][l][0.6]{GPS antenna}
\psfrag{H}[c][c][0.6]{Computer}
\psfrag{I}[c][c][0.6]{Router}
\psfrag{K}[c][c][0.6]{Air part, loaded on the UAV}
\psfrag{L}[c][c][0.6]{Ground part}
\psfrag{U}[c][c][0.6]{on the ground}
\psfrag{N}[l][l][0.6]{Transmission cable}
\psfrag{O}[l][l][0.6]{USB cable}
\psfrag{P}[l][l][0.6]{Ethernet cable}
\psfrag{P}[l][l][0.6]{Ethernet cable}
\psfrag{M}[l][l][0.6]{connect through local area network}
\psfrag{T}[c][c][0.6]{\red Controlling part}
\psfrag{S}[c][c][0.6]{\red Receiver part}
\psfrag{R}[r][r][0.4]{Ref out}
\includegraphics[width=0.48\textwidth]{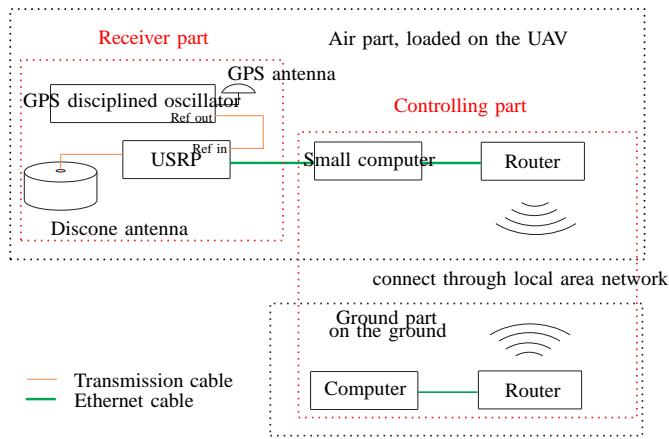}
\caption{A diagram of the equipment used in the measurement campaign.\label{fig:measurement_equipment}}
\end{figure}

\begin{figure}
\centering
\psfrag{A}[c][c][0.8][90]{48 cm}
\psfrag{B}[c][c][0.8]{95 cm}
\psfrag{C}[c][c][0.8]{Packaged discone antenna}
\includegraphics[width=0.4\textwidth]{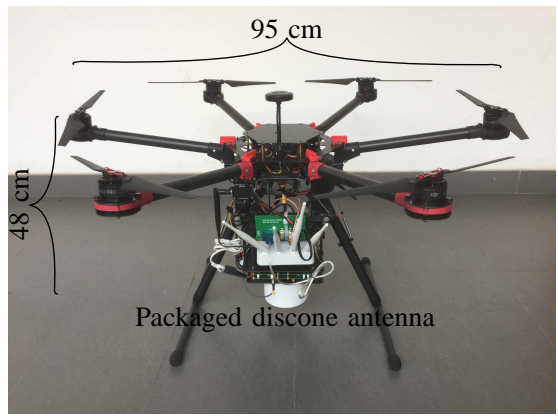}
\caption{The six-wings UAV used in the measurement campaign.\label{fig:uav}}
\end{figure}

Fig.\ \ref{fig:measurement_equipment} illustrates a diagram of the equipment for recording the LTE downlink signals in the measurement campaign. It consists of two parts, i.e., the air part and the ground part. The air part was loaded on the UAV as illustrated in Fig.\ \ref{fig:uav}. It contains the following components: a quasi-omnidirectional packaged discone antenna that works in 1-8 GHz frequency band, a Universal Software Radio Peripheral (USRP) device of type N210 \cite{USRP} which can be programmed to receive real-time signals at specific carrier frequency and with specific sampling rate (or bandwidth), an accurate 10\,MHz reference generated by a Global Positioning System (GPS) disciplined oscillator and then provided to the USRP device, a small computer base unit that controls the USRP device and stores the received data, and a commercial wireless fidelity (WiFi) router. The ground part contains a laptop computer and another commercial WiFi router. By establishing a local area network using the two commercial routers, it is easy and prompt to control the equipment onboard UAV to receive LTE signals via the ground laptop. In addition, the routers worked at the frequency band of 2.4 GHz causing no interference to the LTE signals.

%In practice, it is challenge and cumbersome to connect accessaries, e.g., screen, mouse and keyboard to the computer base unit on the UAV. Thus, we establish the local area network by using the two commercial routers, so that it is easy and prompt to control the equipment onboard the UAV to receive LTE signals by using the ground part.

\subsection{Scenario and specifications}\label{section:scenario}
\begin{figure}
\centering
\psfrag{A}[c][c][0.6]{\white 1}
\psfrag{B}[c][c][0.6]{\white 2}
\psfrag{C}[c][c][0.6]{\white 3}
\psfrag{D}[c][c][0.6]{\white 4}
\psfrag{E}[c][c][0.6]{\white 5}
\psfrag{F}[c][c][0.6]{\white BS}
\psfrag{G}[c][c][0.7]{\white Building A}
\psfrag{H}[c][c][0.7]{\white Building B}
\includegraphics[width=0.45\textwidth]{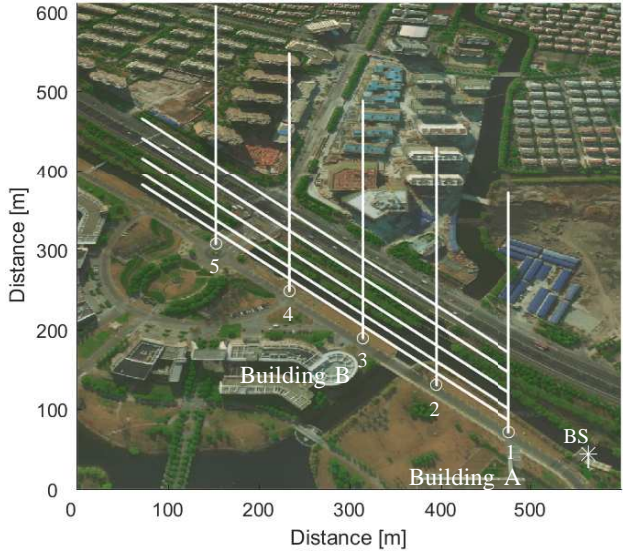}
\caption{Satellite view of the measurement scenario and flight routes.\label{fig:satellite_view}}
\end{figure}

\begin{figure}
\centering
\psfrag{U}[c][c][0.8]{UAV}
\psfrag{S}[c][c][0.8]{BS}
\includegraphics[width=0.4\textwidth]{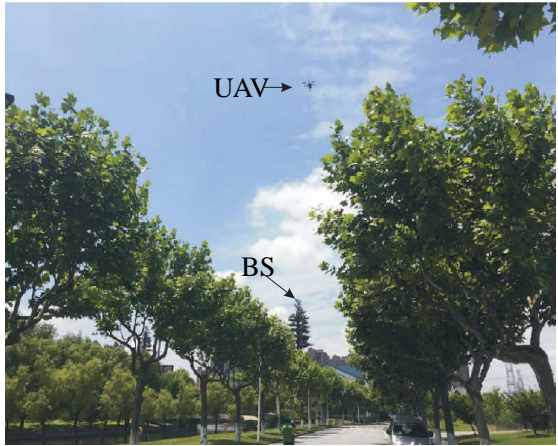}
\caption{A photo taken during the measurement.\label{fig:scenariophoto}}
\end{figure}

\begin{figure}
\centering
\includegraphics[width=0.33\textwidth]{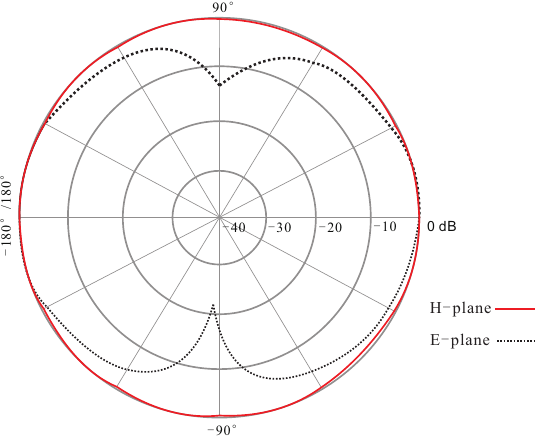}
\caption{The radiation pattern of the discone antenna measured at 2.585 GHz.\label{antenna pattern}}
\end{figure}

The measurement was conducted in a suburban scenario in Jiading Campus, Tongji University, Shanghai. Figs.\ \ref{fig:satellite_view} and\ \ref{fig:scenariophoto} illustrate the satellite view of the measurement scenario and a photo taken during the measurement, respectively. The scenario is characterized by trees, rivers, buildings with four to seven stories etc. The LTE BS is located at the asterisk in Fig.\ \ref{fig:satellite_view} with a height about 20 meters. Five horizontal round-trip flights at the heights of 15, 30, 50, 75, and 100 meters were conducted respectively as indicated by the horizontal lines in Fig.\ \ref{fig:satellite_view}. The round-trip distance was 1000 meters ($500\times2$) and the flying speed was about 5.6 m/s\footnote{The flying speed has a constant value of about 6 m/s except for the phases of starting the movement, braking or changing the direction that occur at the edges of the horizontal lines. Taking into account all the phases, the average speed value is 5.6 m/s.}. We also conducted five vertical flights as indicated by the vertical lines in Fig.\ \ref{fig:satellite_view}. The UAV ascended from the ground to the height of 300 meters at positions 1 to 5, respectively, with a speed of 2.5 m/s. Furthermore, the horizontal distances from these positions to the BS antenna were 100, 200, 300, 400, and 500 meters, respectively. The downlink signals transmitted by the LTE BS were acquired at the carrier frequency of 2.585 GHz and with a complex sampling rate of 25 MHz, and the signal bandwidth was 18 MHz. The radiation pattern of the receiver antenna, i.e., the quasi-omnidirectional discone antenna onboard the UAV was measured at the carrier frequency and is illustrated in Fig.\ \ref{antenna pattern}. It can be observed from Fig.\ \ref{antenna pattern} that the antenna gain changes slightly in the whole azimuth and the elevation between $-60^\circ$ and $60^\circ$. The antenna is chosen so that the radiation pattern is transparent to the channel characteristics.

\subsection{Raw data processing}\label{section:rawdataprocessing}
In the LTE standard, there are two types of frame structure, i.e., frequency division duplex (FDD) and time division duplex (TDD) frame structures. Generally, time duration of one frame is 10 ms which consists of 20 time slots. One slot contains 6 or 7 symbols, and the number depends on the cyclic prefix (CP) length. For example, in the suburban scenario where the delay spread is not so large, short CP is more likely to be applied to mitigate the inter symbol interference, which results in 7 symbols per-slot. Primary synchronization signal (PSS), secondary synchronization signal (SSS) and cell-specific reference signal (CRS) are transmitted on specific symbols in every downlink frame. The exact symbols on which the three signals are transmitted are different, however fixed in FDD and TDD frame structures, respectively. Readers may refer to \cite{ltephystandard} for details. In the post processing of the acquired data (done offline), we exploit the PSS and SSS to accomplish the time synchronization, i.e., to determine the beginning time of a frame in the received raw data and exploit the CRS to extract the CIRs. The considered procedures are detailed as follows.

\emph{Step 1. Filtering:} The raw data received with 25 MHz effective bandwidth at the carrier frequency is firstly low-pass filtered to obtain the 18 MHz baseband signals $r(t)$.

\emph{Step 2. Primary synchronization:} PSS symbol repeats with time interval of half a frame. The location of it in the raw data can be detected by solving the maximization problem:
\begin{align}
(\hat{t}_0,\hat{i}) = \arg \max\limits_{t_0,i}\biggl|\int_{0}^{T}\!\!\!\!r(t)p_{i}^*(t-t_0)\mathrm{d}t\biggr|^2,i= 0,1,2,\label{eq:findinggroupnumber}
\end{align}
where $p_{i}(t)$, $i\in[0,1,2]$ consists of the three kinds of PSS symbol without CP that are generated by the inverse Fourier transform of the corresponding orthogonal frequency division modulation (OFDM) resource elements as specified in \cite{ltephystandard}, $T = 5$ ms is the half frame duration, and $(\cdot)^*$ denotes the complex conjugate of given argument. Moreover, for realistic channels with noise, multipath fading, shadow fading etc. {\red that may cause the temporarily loss of signals, multiple neighboring half-frames can be exploited as}
\begin{align}
(\hat{t}_0,\hat{i}) = \arg \max\limits_{t_0,i}\frac{1}{N}\sum_{n=0}^{N-1}\biggl|\int_{0}^{T}\!\!\!\!r(t+nT)p_{i}^*(t-t_0)\mathrm{d}t\biggr|^2 \label{eq:findinggroupnumber1}
\end{align} {\red to improve the robustness of the detection for $(\hat{t}_0,\hat{i})$,} where $N$ denotes the the number of considered half-frames. Multiple cells with different PSSs transmitted can be identified by considering the dominant peaks.

\emph{Step 3. Secondary synchronization:} One frame contains two SSS symbols with time interval of a half frame. However, they are different with an index difference of 168, i.e., the index of the SSS symbol in the second half frame is that in the first half frame plus 168. SSS symbol and CP mode can be determined by solving the following maximization problem
\begin{small}
\begin{align}
 (\hat{j},\hat{k}) = \arg \max\limits_{j,k}\frac{1}{N}\sum_{n=0}^{N-1}\biggl|\int_{\hat{t}_0+t_j}^{\hat{t}_0+t_j+T_s}r(t+2nT)s_{k}^*(t-\hat{t}_0-t_j)\mathrm{d}t\biggr|^2\;\;\label{eq:detectingj}
\end{align}
\end{small}where $T_s = 1/15$ ms is the duration of one SSS symbol without CP, $t_j, j=[0,1]$ correspond to the time intervals between neighboring SSS symbol and PSS symbol in the same half frame for normal and extended CPs respectively, and $s_{k},k\in[0,1,\dots,335]$ represent the 336 kinds of SSS symbols corresponding to the detected PSS index $\hat{i}$. The physical cell identity (PCI) $N_{id}^{cell}$ can be calculated as
\begin{align}
N_{id}^{cell} = \hat{i} + 3 \times (\hat{k} \bmod 168) \label{eq:pcical}
\end{align}
where ($\cdot$ $\bmod$ $\cdot$) denotes the remainder of the former argument divided by the latter argument. Moreover, $\hat{k}\in[0,1,\dots, 167]$ means that the detected SSS is in the first half frame, otherwise the second half frame.

\emph{Step 4. CIR extraction:} The CRS symbols that span the whole baseband bandwidth in frequency domain are exploited to extract CIRs as
\begin{align}
h(\tau,t_c) = \mathscr{F}^{-1}{\red \bigl(}\frac{c_{r}(f,t_c)}{c_s(f,t_c)}{\red \bigr)} \label{eq:cirextraction}
\end{align}
where $c_{r}(f,t_c)$ is the representation in frequency domain of the received CRS symbol at time $t_c$, $c_s(f,t_c)$ is the representation of the correspondingly sent CRS symbol that is generated referring to \cite{ltephystandard} with the information obtained in previous steps, such as the beginning location of a frame, PCI and CP mode, and $\mathscr{F}^{-1}(\cdot)$ denotes the inverse Fourier transform of given argument. The CIR output rate is 200 per second, i.e., the time interval between neighboring CIRs is 5 ms. Furthermore, we denote $h(\tau,t_c)$ as $h(\tau,t)$ in the sequel for notation convenience.

\begin{figure}
\psfrag{B}[l][l][0.6]{SAGE estimation results}
\psfrag{C}[l][l][0.6]{Noise floor}
\psfrag{A}[l][l][0.6]{Empirical PDP}
\includegraphics[width=0.47\textwidth]{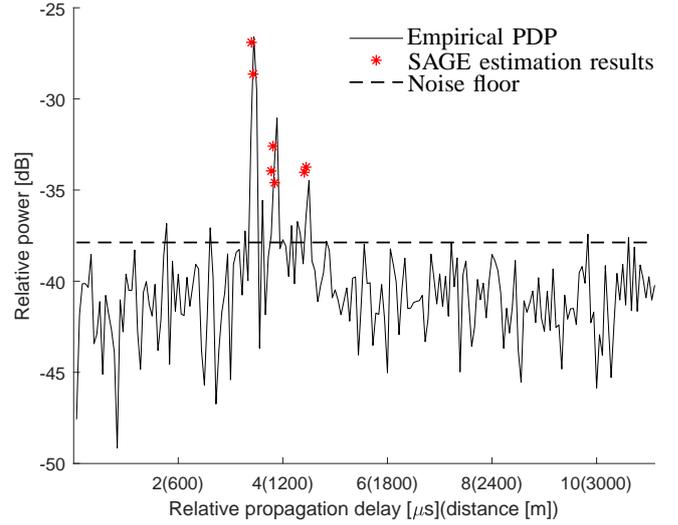}
\caption{An example PDP and the corresponding SAGE estimation results.\label{example_pdp}}
\end{figure}

By exploiting the CIR results, a high resolution algorithm is derived based on the SAGE principle to estimate parameters, i.e., delays, Doppler frequencies and complex amplitudes of MPCs. In the underlying SAGE algorithm, the generic model of the CIR is formulated as
\begin{align}
h(\tau,t)=\sum_{\ell=1}^{L(t)} \alpha_{\ell}(t)\delta(\tau-\tau_{\ell}(t))\exp\biggr\{j2\pi \int_0^t \nu_\ell(t) t \biggr\} + n(\tau,t)\label{eq:spreadfunction}
\end{align}
where $L(t)$ is the total number of paths that exist in the channel snapshot at time $t$, $\alpha_\ell(t)$, $\tau_{\ell}(t)$, and $\nu_\ell(t)$ represent the complex amplitude, delay and Doppler frequency of the $\ell$th path respectively, $\delta(\cdot)$ denotes the Dirac delta function, and $n(\tau,t)$ represents the noise. Readers are referred to \cite{Fleury1999Channel} for the details of the SAGE algorithm. {\red It has been practically concluded in [51] that the SAGE algorithm can resolve two paths with $\Delta \tau \gtrapprox \frac{1}{5B}$, where $\Delta \tau$ and $B$ denote the relative delay between two paths and the signal bandwidth, respectively.} In our implementation, we consider $10$ consecutive CIRs, i.e., a time duration of 50 ms as one snapshot for the channel during which the multipath parameters are observed to be constant. The SAGE algorithm is firstly applied with 30 paths to pre-estimate the MPC parameters, and $L(t)$ is then chosen as the index of the path whose amplitude falls below the noise level. Then the SAGE algorithm is applied again with path number $L(t)$ and 15 iterations to fully extract the components. The parameters estimated are $\bsTheta=[\alpha_{\ell}(t), \tau_{\ell}(t), \nu_{\ell}(t); \ell=1,\dots,L(t), t=t_1,\dots,t_M]$ with $M$ being the number of snapshots. {\red As an example, Fig. \ref{example_pdp} illustrates a power delay profile (PDP) for the horizontal flight at the height of 75 meters (see the concatenated PDPs (CPDPs) in Fig. \ref{hozizontal_flight_75m} ). The noise floor is estimated by adding 3 dB to the mean power of tail samples in the PDP. It can be observed that the SAGE algorithm is capable to estimate the MPCs.}

\section{Channel characterization}\label{channel_characterization}
In this section, we present the measurement results for the horizontal flights and the vertical flights, which include the CPDPs and the corresponding MPC parameters estimated by using the SAGE algorithm. These results shed lights on the propagation mechanisms of the UAV channels.

\subsection{Horizontal flights}

\begin{figure}
%\psfrag{segment index}[c][c][0.8][90]{Time [s]}
%\psfrag{Delay [s]}[c][c][0.8]{Relative propagation delay [$\mu$s]}
%\psfrag{Power [dBm]}[c][c][0.8][90]{Received power [dbw]}

\psfrag{D}[c][c][0.8]{Doppler frequency [Hz]}
\psfrag{B}[c][c][0.8]{Relative propagation delay [s]}
\psfrag{C}[c][c][0.8][90]{Relative power [dB]}
\psfrag{A}[c][c][0.8][90]{Time [s]}
\psfrag{E}[c][c][0.8][90]{Time [s]}
\psfrag{F}[c][c][0.8][]{Time [s]}
\subfigure[CPDPs\label{hozizontal_flight_75m}]{\includegraphics[width=0.47\textwidth]{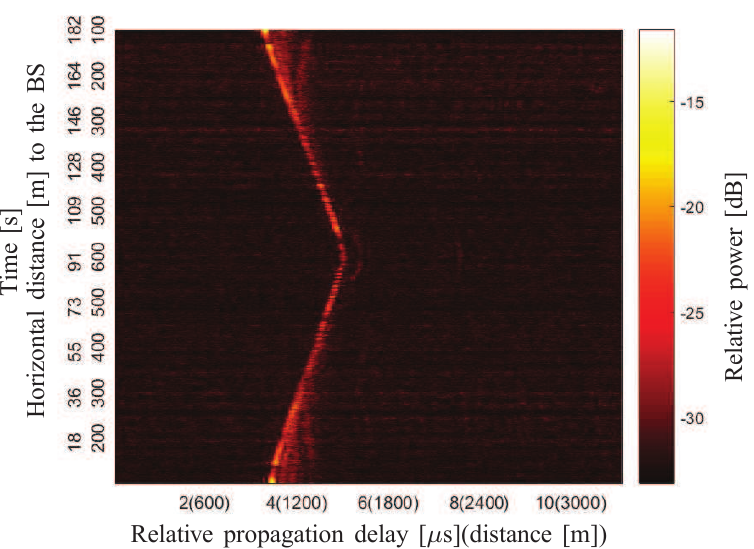}}%{horizontal_flight_75m_151317}}%
%\rput[tr]{0}(-1.8,0.2){ \footnotesize Relative propagation delay [$\mu$s](distance [m])}
%\rput[tr]{90}(-8.2,5.3){\footnotesize Horizontal distance [m] to the BS}
%\rput[tr]{90}(-0.5,4.1){\footnotesize Relative power [dB]}
%\rput[tr]{90}(-8.5,3.7){\footnotesize Time [s]}

\centering
\subfigure[SAGE estimation results in delay domain\label{sage_delay_75m}]{\includegraphics[width=0.45\textwidth]{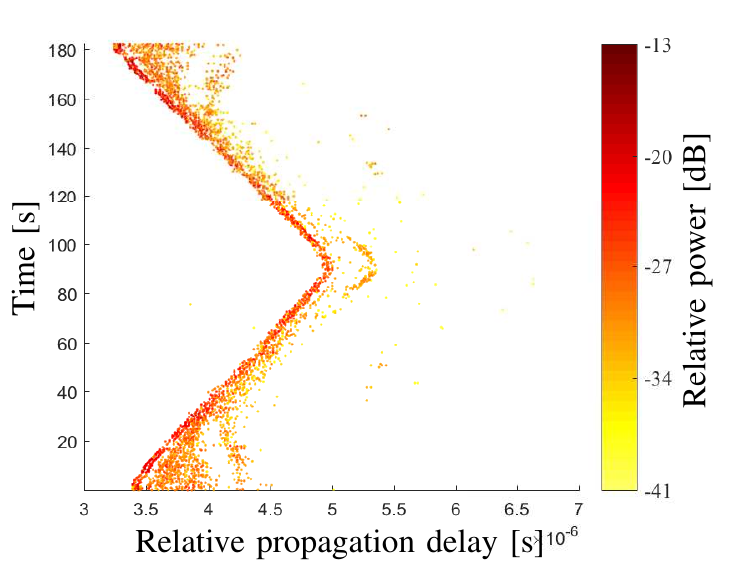}}
\subfigure[SAGE estimation results in Doppler frequency domain\label{sage_doppler_75m}]{\includegraphics[width=0.45\textwidth]{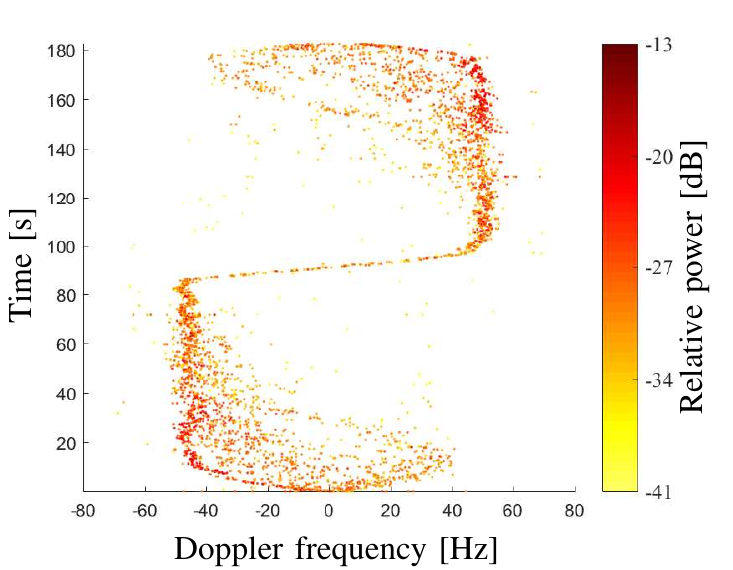}}
\caption{CPDPs of a horizontal round-trip flight at the height of {75} meters and corresponding SAGE estimation results.\label{overallh}}
\end{figure}

Fig.\ \ref{hozizontal_flight_75m} illustrates the CPDPs of a horizontal round-trip flight at the height of {{75}} meters. The horizontal and vertical axes represent the propagation delay and the measurement time, respectively, and the color denotes the received power. It is noteworthy that we cannot obtain the absolute propagation delay in the passive sounding measurement, since the time (or delay) synchronization elaborated in Sect. \ref{section:rawdataprocessing} is obtained relative to the path with the highest power. Thus we use the term ``relative propagation delay''. Nevertheless, the link distance is easily calculated by exploiting the geometry information so that the absolute propagation distances of multiple paths can be retrieved. Similarly, we use ``relative power'' because the transmitted power of BS antenna is unknown.

It can be observed from Fig.\ \ref{hozizontal_flight_75m} that the variation of the path trajectory is consistent with the flight route, i.e., the propagation distance increased first and then decreased. It took about 90 seconds for the UAV to fly a single journey of 500 meters. The symmetry of the channel is observed as expected. In order to obtain further insights into the channel, the SAGE estimation results in delay and Doppler frequency domains are illustrated respectively in Figs.\ \ref{sage_delay_75m} and \ref{sage_doppler_75m} for the CPDPs in Fig.\ \ref{hozizontal_flight_75m}. It is obvious that the Doppler frequency trajectory is also in accordance with the flight route. The absolute Doppler frequency of the {\red LoS} path firstly increased as the UAV speed increased from 0. When the UAV speed reached its maximum and the angle between the speed and the LoS path was almost -180$^\circ$ with the horizontal distance increasing, the absolute LoS Doppler frequency remained almost constant for a certain time. Then the UAV decreased its speed for turning around, which results in decreasing absolute LoS Doppler frequency to 0\,Hz. Furthermore, rich MPCs that spread in delay and Doppler frequency domains can be observed from 0 to 35 seconds. We postulate that it is due to the signals interacting with the local scatterers such as buildings A and B, trees, and the ground (see Fig.\ \ref{fig:satellite_view}). For example, by examining the orientations of the facades of buildings A and B and the BS location, it can be inferred that the signals interacting with the two buildings can be effectively received by the UAV in the first 200 meters of the horizontal route. After the UAV passed by this area, the number of MPCs decreases due to the disappearance of these contributions. Basically, the received signal power decreases with the UAV flying away from the BS antenna in the whole journey. In addition, some dark segments exist on the red trajectory in Fig.\ \ref{hozizontal_flight_75m}, which means that deep fading existed intermittently during the flight. It can be inferred that this is caused by the destructive addition of MPCs, e.g., the  LoS path component and the ground reflection path component.

\begin{figure}
	
	\psfrag{D}[c][c][0.8]{Doppler frequency [Hz]}
    \psfrag{B}[c][c][0.8]{Relative propagation delay [s]}
    \psfrag{C}[c][c][0.8][90]{Relative power [dB]}
    \psfrag{A}[c][c][0.8][90]{Time [s]}
    \psfrag{E}[c][c][0.8][90]{Time [s]}
    \psfrag{F}[c][c][0.8][]{Time [s]}
	\subfigure[CPDPs\label{vertical_flight_marker_2}]{\includegraphics[width=0.47\textwidth]{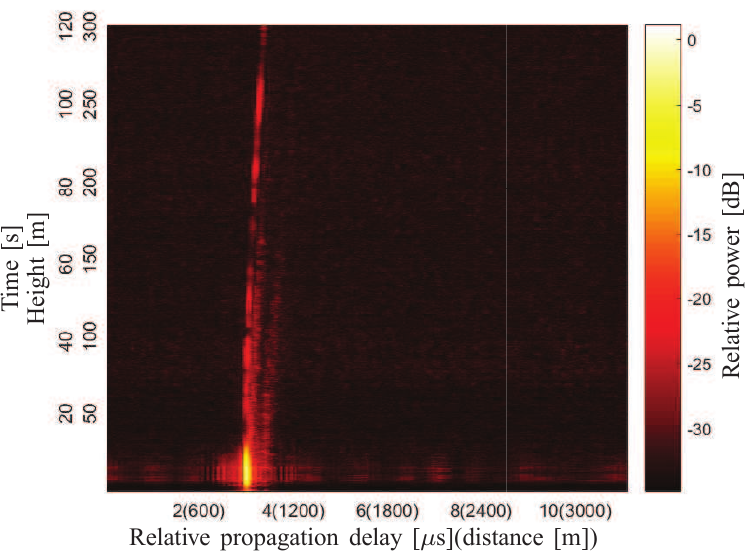}}%{vertical_flight_161053}}%
    %\rput[tr]{0}(6.3,7.3){ \footnotesize Relative propagation delay [$\mu$s](distance [m])}
%    \rput[tr]{90}(-0.1,10.85){\footnotesize Height [m]}
%    \rput[tr]{90}(7.7,11.3){\footnotesize Relative power [dB]}
%    \rput[tr]{90}(-0.4,10.7){\footnotesize Time [s]}

\centering
	\subfigure[SAGE estimation results in delay domain\label{sage_delay_vertical}]{\includegraphics[width=0.45\textwidth]{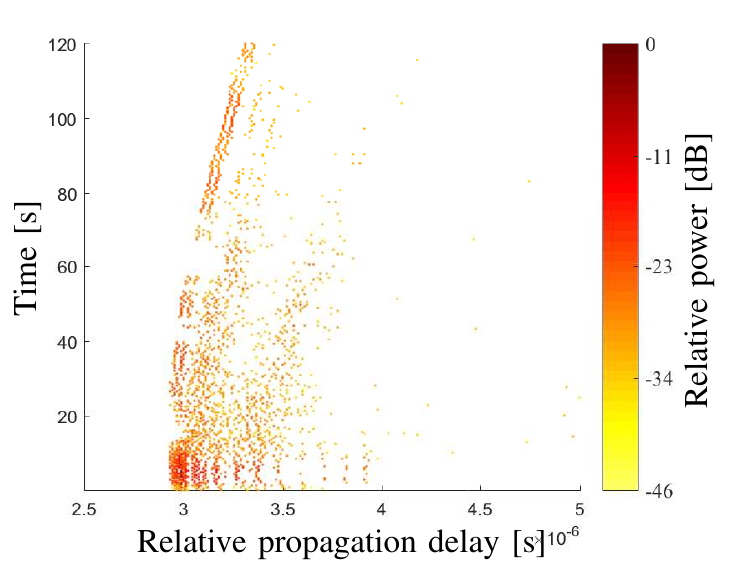}}
	\subfigure[SAGE estimation results in Doppler frequncy domain\label{sage_doppler_vertical}]{\includegraphics[width=0.45\textwidth]{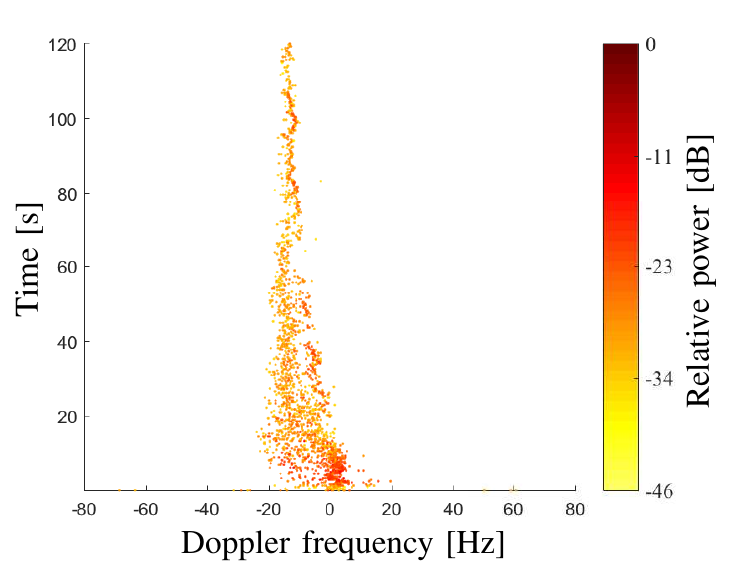}}
	\caption{CPDPs of a vertical flight ascending at ascending position 2 and corresponding SAGE estimation results.\label{overallv}}
\end{figure}

\subsection{Vertical flights}\label{sect:vertical_channel_characterization}
Fig.\ \ref{vertical_flight_marker_2} illustrates an example of CPDPs for the vertical flights. The UAV flew up to 300 meters height in 120 seconds. The descending data was not recorded in order to save the storage since the descending speed was very slow for safety. The SAGE estimation results in delay and Doppler frequency domains are also illustrated in Figs.\ \ref{sage_delay_vertical} and \ref{sage_doppler_vertical}, respectively. It can be observed from Fig.\ \ref{sage_doppler_vertical} that the LoS Doppler frequency varied from positive values to negative ones, which is because of the BS antenna height, i.e., the UAV approached and then left away from the BS in the ascending process. Afterwards, the absolute LoS Doppler frequency approached to the maximum value with the height increasing. The channel in low heights exhibited much more MPCs, and less MPCs were observed with increasing height. We postulate the reasons are as follows. One is that the power of non-LoS (NLoS) MPCs caused by the ground scatterers decayed more significantly, because the propagation distances of them changed more than that of the LoS path with the height increasing. The other one is that some scatterers, e.g., buildings A and B, contributed less to the received signal at the higher height due to the orientations of their facades.
Furthermore, it is interesting to observe in Fig.\ \ref{sage_doppler_vertical} that almost all the NLoS Doppler frequencies locate in the left side of that of LoS path, {\red and negative Doppler frequencies exist near the ground. It can be inferred that the channel includes multiple types of propagation routes, e.g., from Tx to UAV (Tx-UAV), Tx-ground-UAV, Tx-building-UAV, Tx-building-ground-UAV, etc.} Similar explanations can be applied to the NLoS path distributions for the horizontal flight as illustrated in Fig.\ \ref{sage_doppler_75m}. Moreover, intermittent deep fading can also be observed in the vertical flights due to, e.g., the destructive addition of MPCs.% and blockage.

\section{UAV Channel model}\label{channel_model}
One of the most important statistics of the channel in order to design a communication system is the attenuation or, conversely, the channel power. It will affect the most basic aspects of design, such as the required transmit power (for a desired link range) or the modulation and coding scheme, hence the achievable rate. %Note also that the desired link range will impose requirements on the Automatic Gain Control (AGC) specifications. Based on the SAGE estimation results...
Based on the SAGE estimation results, the power of the channel is calculated as
\begin{equation}
P(t)=\sum\limits_{\ell=1}^{L(t)} \left|\alpha_{\ell}(t)\right|^2 \label{eq:powercalculation}
\end{equation} where $\left|\cdot\right|$ denotes the absolute value of the argument.

\begin{figure}
	\centering
	\includegraphics[width=0.47\textwidth]{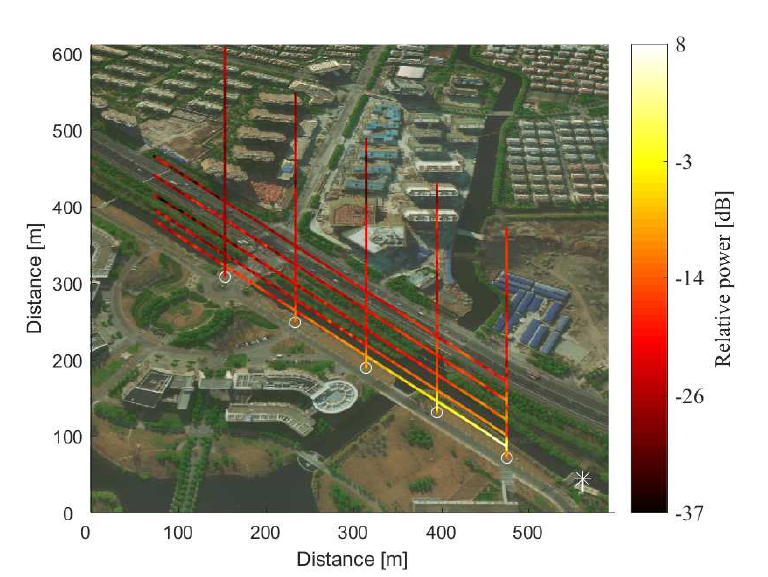}
	\caption{Received signal power for five horizontal and five vertical flights in the satellite map. \label{power_in_map}}
\end{figure}

Fig.\ \ref{power_in_map} illustrates the relative power of received signals for the five horizontal and five vertical flights in the satellite map. It can be clearly observed that the channels for different horizontal or vertical flights are distinctive. Especially the channel power of the horizontal flight at the height of 15 meters is obviously higher than that for all the other flights. We postulated that it is mainly because the BS antenna is tilted towards the ground. To gain further insights into the channel characteristics, path loss, shadow fading, fast fading, root mean square (RMS) delay spread and RMS Doppler frequency spread are analyzed for horizontal flights and vertical flights.

\subsection{Horizontal flights}

\begin{figure}
	\centering
	\psfrag{data1}[l][l][0.7]{Height of 15 m}
	\psfrag{data2}[l][l][0.7]{Height of 30 m}
	\psfrag{data3}[l][l][0.7]{Height of 50 m}
	\psfrag{data4}[l][l][0.7]{Height of 75 m}
	\psfrag{data5}[l][l][0.7]{Height of 100 m}
	\psfrag{distance}[c][c][0.8]{Horizontal distance to Tx [m]}
	\psfrag{power}[c][c][0.8]{Relative power [dB]}
    \includegraphics[width=0.47\textwidth]{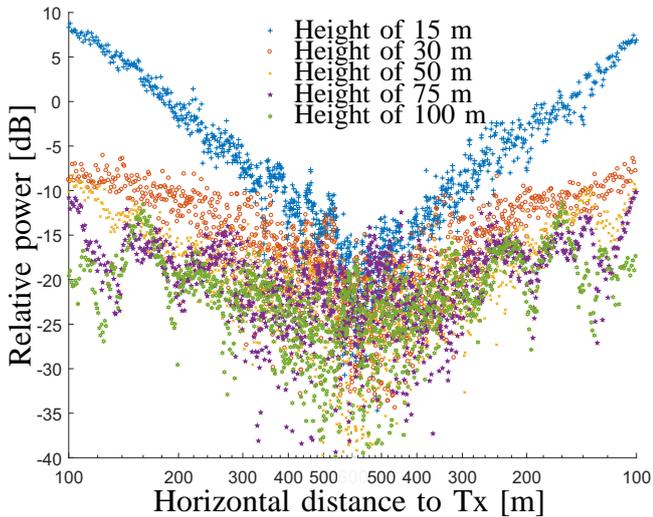}
	\caption{Received power for the five horizontal round-trip flights at the heights of 15 m, 30 m, 50 m, 75 m and 100 m.\label{horizontal_power}}
\end{figure}

Fig.\ \ref{horizontal_power} illustrates the received power for the five horizontal flights at the heights of 15, 30, 50, 75 and 100 meters, respectively.  We have the following preliminary observations from Fig.\ \ref{horizontal_power}: \textit{i)} In general, the received power decreases with the horizontal distance increasing for one horizontal fight. It is straightforward since the link distance gets larger; \textit{ii)}  In general, the power decay slope with respect to horizontal distance becomes lower at a higher height;  \textit{iii)} Fast fading observed is noticeably less severe for 15 m height than for the other cases.
 %at the other four heights is obviously severer than that of 15 meters. The investigations and explanations for the characteristics are as follows.

\subsubsection{Path loss}

\begin{figure}
	\centering
	%\psfrag{distance1}[c][c][0.8]{Horizontal distance [m] at 15 meters' height}
%\psfrag{distance2}[c][c][0.8]{Horizontal distance [m] at 30 meters' height}
%\psfrag{distance3}[c][c][0.8]{Horizontal distance [m] at 50 meters' height}
%\psfrag{distance4}[c][c][0.8]{Horizontal distance [m] at 75 meters' height}
%\psfrag{distance5}[c][c][0.8]{Horizontal distance [m] at 100 meters' height}
%	\psfrag{power}[c][c][0.8]{Power [dBm]}
%	\psfrag{empirical}[l][l][0.8]{Empirical}
%    \psfrag{empirical2}[l][l][0.8]{Empirical}
%    \psfrag{empirical3}[l][l][0.8]{Empirical}
%    \psfrag{empirical4}[l][l][0.8]{Empirical}
%    \psfrag{empirical5}[l][l][0.8]{Empirical}
%	\psfrag{fitted}[l][l][0.8]{Fitted}
    \psfrag{distance}[c][c][0.8]{Horizontal distance [m]}
	\psfrag{power}[c][c][0.8]{Relative power [dB]}
	\psfrag{data1}[l][l][0.7]{Empirical, height of 15 m}
    \psfrag{data2}[l][l][0.7]{Fitted}
    \psfrag{data3}[l][l][0.7]{Empirical, height of 30 m}
    \psfrag{data4}[l][l][0.7]{Fitted}
   \psfrag{data5}[l][l][0.7]{Empirical, height of 50 m}
   \psfrag{data6}[l][l][0.7]{Fitted}
     \psfrag{data7}[l][l][0.7]{Empirical, height of 75 m}
    \psfrag{data8}[l][l][0.7]{Fitted}
   \psfrag{data9}[l][l][0.7]{Empirical, height of 100 m}
   \psfrag{data10}[l][l][0.7]{Fitted}
	\includegraphics[width=0.47\textwidth]{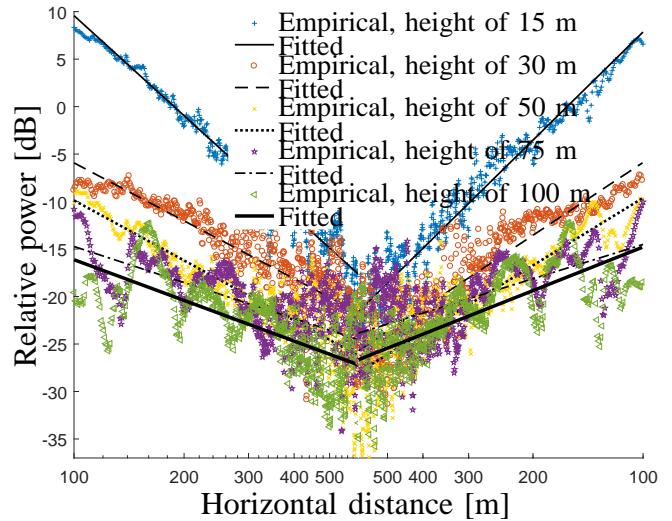}
	\caption{Power fittings for the five horizontal flights. \label{path_loss_fitting}}
\end{figure}

We propose a modified close-in free path loss model by exploiting the horizontal flights as
%\begin{align}
%P_\mathrm{L}[dB]  = P_{\mathrm{L},h}(d_0) {+}10\gamma_{h} \cdot \log_{10}(\frac{d}{d_0})+ X_h \label{eq:splmodel}% + 10\gamma_{h} \cdot \log_{10}(\frac{d_h}{d_0})  \label{eq:splmodel}
%\end{align} where $d_0$ is the reference distance (which is set as 1 m in our case), $P_{\mathrm{L},h}(d_0)$ is the path loss at $d_0$\footnote{In our case, $P_{\mathrm{L},h}(d_0)$ are relative values, thus we do not present them in the analysis in the sequel.}, $d$ is the horizontal distance, $\gamma_{h}$ represents the path loss exponent which is related to the height $h$, and $X_h$ denotes the shadow fading.
{\red \begin{align}
P_\mathrm{L}[dB]  = 10\gamma_{h} \cdot \log_{10}(d)+ X_h + b_h \label{eq:splmodel}% + 10\gamma_{h} \cdot \log_{10}(\frac{d_h}{d_0})  \label{eq:splmodel}
\end{align} where $d$ is the horizontal distance, $\gamma_{h}$ represents the path loss exponent which is related to the height $h$, $X_h$ denotes the shadow fading, and $b_h$ represents the intercept\footnote{In our case, $b_h$ are relative values, thus we do not present them in the analysis in the sequel}.}
Fig.\ \ref{path_loss_fitting} illustrates the power curve fitting for the five horizontal flights. A sliding/overlapped window of 20 wavelengths \cite{1623289} is applied to removing the fast fading before the fittings, and the smoothed power is also illustrated in Fig.\ \ref{path_loss_fitting}. Table\ \ref{tab:ple} presents the path loss exponents obtained for the five horizontal flights. It can be observed that generally $\gamma_h$ is negatively correlated with height $h$. We postulate the reasons are as follows. \textit{i)} At a higher height, the link distance is less sensitive to the horizontal distance. However, since the height is not so large when compared to the horizontal distance in log-scale, this effect can be almost negligible. {\blue \textit{ii)} Usually, the beam of the LTE BS antenna is empirically downward covering a certain range on the ground. The UAV at low heights is more likely to experience significant power change when it flies from the main beam towards non-dominant radiation area of BS antenna.} We use the generic model
\begin{align}
\gamma_h = a\cdot h+b+c
\end{align}
with $a$ and $b$ being constants and $c$ being a zero-mean normally distributed variable to model the path loss exponents. $a$ and $b$ are calculated as -0.02 and 3.42, respectively, and the standard deviation of $c$ is calculated as 0.48.

\begin{table}
\centering
%\psfrag{Power angular spectra, [dB]}[l][l][0.6]{}
\caption{Path loss exponents for horizontal flights}
\scalebox{0.9}{
\begin{tabular}{cccccc}
\hline
%\multicolumn{4}{c}{{{{\textit{Measurement specifications}}}}}\\\hline
Height [m] &  15 & 30 & 50 & 75 & 100     \\
Outbound $\gamma_h$ &  3.50 & 2.04 & 2.12 & 1.34 & 1.62     \\
Inbound $\gamma_h$ &  3.78 & 2.56 & 2.42 & 1.27 & 1.73     \\
Mean $\gamma_h$ &  3.64 & 2.30 & 2.28 & 1.31 & 1.67     \\ \hline
%Tx antenna height & $1.5$ meters & Rx antenna height & $1.9$ meters\\ \hline
\end{tabular}}
\label{tab:ple}
\end{table}

\subsubsection{Shadow fading}
\begin{figure}
	\centering
\psfrag{data1}[l][l][0.7]{Height of 15 m}
\psfrag{data2}[l][l][0.7]{Height of 30 m}
\psfrag{data3}[l][l][0.7]{Height of 50 m}
\psfrag{data4}[l][l][0.7]{Height of 75 m}
\psfrag{data5}[l][l][0.7]{Height of 100 m}
\psfrag{distance}[c][c][0.8]{Horizontal distance [m]}
\psfrag{shadow}[c][c][0.8]{$X_h$ [dB]}
\subfigure[]{\includegraphics[width=0.47\textwidth]{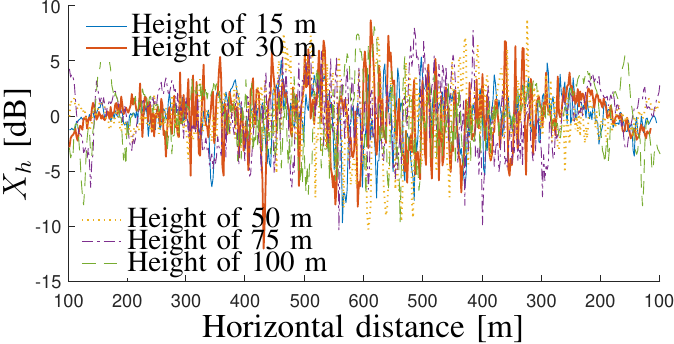}\label{horizontal_shadowing}}
\psfrag{shadow data}[l][l][0.7]{Empirical, 15 to 100 m}
\psfrag{fit 10}[l][l][0.7]{${\mathcal{N}(0,2.7)}$}
\psfrag{Cumulative probability}[c][c][0.8]{CDF $P(X_h [\text{dB}] < abscissa)$}
\psfrag{Data}[c][c][0.8]{$X_h$ [dB]}
\subfigure[]{\includegraphics[width=0.47\textwidth]{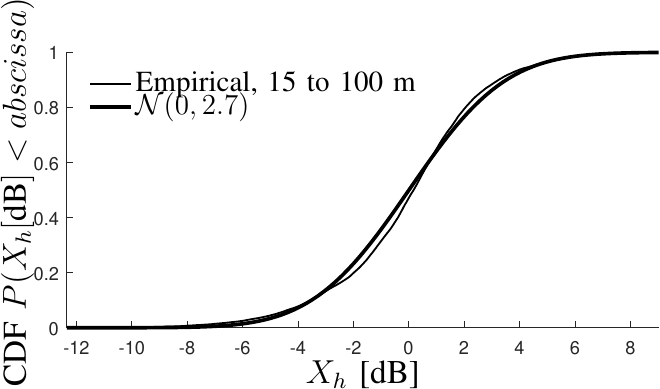}\label{shadow_fading_cdfs}}
\caption{(a) Shadow fading variation for the five horizontal flights. (b) CDF of shadow fading for the five horizontal flights.}
\end{figure}

Shadow fading $X_h$ for the horizontal flights is calculated by subtracting path loss from the smoothed power. Fig.\ \ref{horizontal_shadowing} illustrates the shadow fading variation for the five horizontal flights at different heights. {\red It can be observed that generally the shadow fading deviation increases when either the height or horizontal distance increases. It is reasonable since the randomly addition of LoS path and ground-surface scattered paths dominates the channel with ground objects contributing less. The BS antenna pattern may also affect the shadow fading, however, we do not consider it is the main reason, since when the UAV experienced similar sections of the BS antenna radiation pattern during the four horizontal flights with heights above the BS antenna height, the shadow fading patterns are different (or not similar).} %{\red It can be observed that generally the shadow fading deviation increases when either the height or horizontal distance increases. It is reasonable since the randomly addition of LoS path and ground-surface scattered paths dominates the channel with weaker contributions from the ground objects. Although the BS antenna pattern may also affect the shadow fading, we do not consider it is the main reason. Since with UAV experienced almost the same pattern area, the shadow fading of the four horizontal flights above 20 m (BS height) is not the same.}
%It can be observed that generally the absolute shadow fading increases with the horizontal distance increasing.One possible reason is that when the UAV is far away from the BS, the LoS path is more likely to be blocked; the other reason is that when the UAV flies away, it is more likely to experience the sidelobes of the BS antenna.
Moreover, % it can be observed that the shadow fading for the five horizontal flights are similar. Actually,
calculations show that the maximum difference among their standard deviations is less than 0.5 dB. For model simplicity, we use one CDF to model them.
Note that in the sequel, similar CDFs of the same parameters will also be combined for simplicity.
Fig.\ \ref{shadow_fading_cdfs} illustrates the cumulative distribution function (CDF) of the shadow fading for the five horizontal flights. Normal distribution\footnote{In the representation $\mathcal{N}(\mu,\sigma)$, $\mu$ and $\sigma$ represent respectively the expectation and standard deviation of target variable.} ${\mathcal{N}(0,2.7)}$ is found to best fit the empirical distribution. Furthermore, the correlation coefficient between the absolute shadow fading and the horizontal distance is calculated as 0.36. The dependency of shadow fading on horizontal distance can be modeled by considering the horizontal distance as a uniform distributed random variable and the shadow fading as a log-normal distributed random variable with the correlation coefficient applied. In addition, the correlation coefficient $\rho$ between two random variables $x$ and $y$ is calculated as
\begin{align}
\rho(x,y) = \frac{\text{Cov}(x,y)}{\sqrt{\text{Var}(x)\text{Var}(y)}}
\end{align} where $\text{Cov}(\cdot,\cdot)$ and $\text{Var}(\cdot)$ denotes the covariance and variance of the argument(s), respectively.

\subsubsection{Ricean K factor}\label{sect:horizontal_k}

\begin{figure}
	\centering
\psfrag{data1}[l][l][0.7]{Height of 15 m}
\psfrag{data2}[l][l][0.7]{Height of 30 m}
\psfrag{data3}[l][l][0.7]{Height of 50 m}
\psfrag{data4}[l][l][0.7]{Height of 75 m}
\psfrag{data5}[l][l][0.7]{Height of 100 m}
\psfrag{distance}[c][c][0.8]{Horizontal distance [m]}
\psfrag{kfactor}[c][c][0.8]{$K_h$ [dB]}
\subfigure[]{\includegraphics[width=0.47\textwidth]{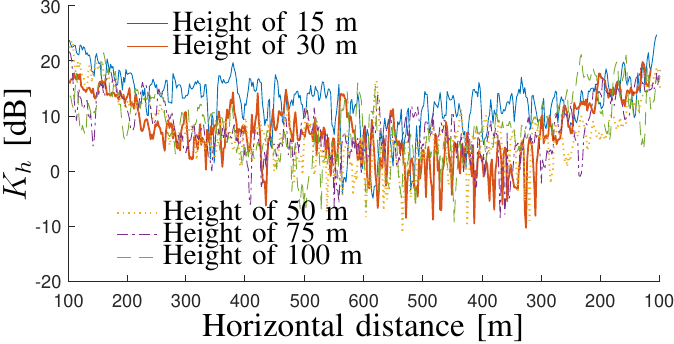}\label{fast_fading_variation}}

\psfrag{k_15}[l][l][0.7]{Empirical $K$, 15 m}
\psfrag{k15to100}[l][l][0.7]{Empirical $K$, 30 to 100 m}
\psfrag{fit_1}[l][l][0.7]{${\mathcal{N}(12.6,5.1)}$}
\psfrag{fit_2}[l][l][0.7]{${\mathcal{N}(7.6,5.6)}$}
\psfrag{Cumulative probability}[c][c][0.8]{CDF $P(K_h[\text{dB}] < abscissa)$}
\psfrag{Data}[c][c][0.8]{$K_h$ [dB]}
\subfigure[]{\includegraphics[width=0.47\textwidth]{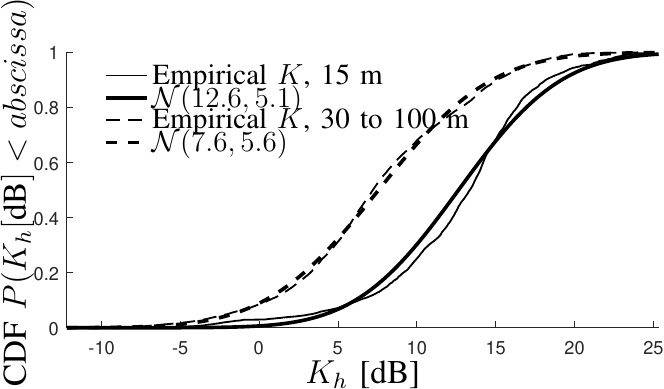}\label{fast_fading_cdfs}}
\caption{(a) K factor vs. horizontal distance for the five horizontal flights. (b) CDFs of K factors for the five horizontal flights.}
\end{figure}

The Ricean K-factor characterizes the statistical distribution of the received signal amplitude. It is the ratio of the power in the LoS component or dominant component to the power in the NLoS or the other multipath components \cite{7489014}. Based on the power $P$, we use the classical moment-based method proposed in \cite{greensteinmethod} to calculate the K factors. Fig.\ \ref{fast_fading_variation} illustrates the K factor vs. horizontal distance at different heights. It can be observed from Fig.\ \ref{fast_fading_variation} that generally the K factor becomes smaller with the horizontal distance increasing, although the channel is more LoS-alike with less NLoS paths, as illustrated in Fig.\ \ref{sage_delay_75m}. {\blue Our conjecture is that with horizontal distance increasing the power of LoS path decays more rapidly due to the fact that the UAV flies outwards the main beam of BS antenna, {\red while the NLoS paths are almost always caused by the interactions between the main beam of BS and the ground objects and surface.}} %The LoS path is also more likely to be blocked when the UAV is far away, which can result in lower K factor.
It is interesting to observe in Fig.\ \ref{fast_fading_variation} that the K-factor at the height of 15 meters is larger than that of the other four heights. {\blue We postulate that it is mainly due to the fact that the main beam of BS antenna can cover the most of the flight route at the height of 15 meters, while for the other four heights, the UAV is out of the main beam most of the time.} Fig.\ \ref{fast_fading_cdfs} illustrates the CDFs for the K factors at different heights. The CDFs of K factors at the heights from 30 to 100 meters are combined as one, since they are very similar. $\mathcal{N}(12.6,5.1)$ and $\mathcal{N}(7.6,5.6)$ are found to best fit the empirical ones. Moreover, the correlation coefficients for the height of 15 meters and for the other four heights are calculated as -0.64 and -0.65, respectively.

\subsubsection{RMS delay spreads and Doppler frequency spreads}
Delay spread and Doppler frequency spread are very important features for the design of higher layers of communication systems. The delay spread is inversely related with the channel coherence bandwidth, hence a high delay spread will affect the cyclic prefix length. High delay spread will lead to inter-symbol interference, requiring the use of more advanced equalization architectures at the receiver. Regarding the Doppler frequency spread, it is inversely related with the channel coherence time. A higher spread will limit the maximum frame size and could be crucial to decide the duplexing method to be used for the communications. It can also limit the maximum bandwidth per subcarrier, since it leads to inter-carrier interference (ICI), hence making it necessary to include ICI cancellation methods at the receiver or more advanced channel equalization techniques.

The RMS delay spread and Doppler frequency spread are calculated as the second-order central moments of the power delay profile and power Doppler frequency profile, respectively. Specifically, by using the SAGE estimation results, the RMS delay spread is calculated as specified in \cite{7489014,Yin2015Propagation}
\begin{equation}
\sigma_{\tau}(t)=\sqrt{\overline{\tau^2(t)}-\bar{\tau}^2(t)}\label{eq:spreadcomputing}
\end{equation}with
\begin{align}
\overline{\tau^2}(t)&=\frac{\sum\limits_{\ell=1}^{L(t)}|\alpha_\ell|^2\tau_\ell^2}{\sum\limits_{\ell=1}^{L(t)}|\alpha_\ell|^2}, \;\; \bar{\tau}(t)=\frac{\sum\limits_{\ell=1}^{L(t)}|\alpha_\ell|^2\tau_\ell}{\sum\limits_{\ell=1}^{L(t)}|\alpha_\ell|^2}.\label{eq:bartau}
\end{align}
 The RMS Doppler frequency spread is calculated similarly by using (\ref{eq:spreadcomputing}) and (\ref{eq:bartau}) with $\tau$ replaced by $\nu$.

\begin{figure}
	\centering
\psfrag{data1}[l][l][0.7]{Height of 15 m}
\psfrag{data2}[l][l][0.7]{Height of 30 m}
\psfrag{data3}[l][l][0.7]{Height of 50 m}
\psfrag{data4}[l][l][0.7]{Height of 75 m}
\psfrag{data5}[l][l][0.7]{Height of 100 m}
\psfrag{distance}[c][c][0.8]{Horizontal distance [m]}
\psfrag{delayspread}[c][c][0.8]{$\log_{10}(\sigma_{\tau,h})[\text{s}]$}
\subfigure[]{\includegraphics[width=0.47\textwidth]{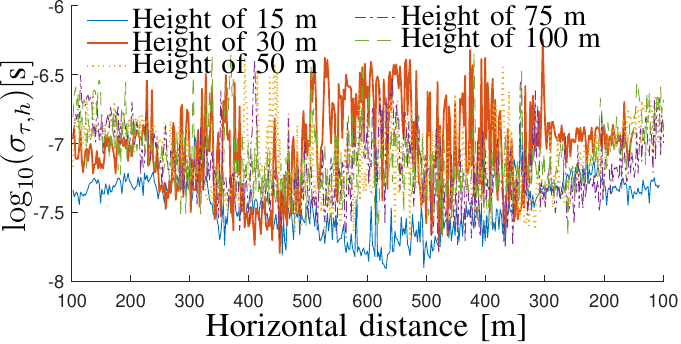}\label{horizontal_delay_spreads}}
\psfrag{delay_spreads_15}[l][l][0.7]{Empirical, 15 m}
\psfrag{delayspreads30to100}[l][l][0.7]{Empirical, 30 to 100 m}
\psfrag{fit_1}[l][l][0.7]{${\mathcal{N}(-7.4,0.2)}$}
\psfrag{fit_2}[l][l][0.7]{${\mathcal{N}(-7.1,0.3)}$}
\psfrag{meanvar}[l][l][0.7]{-7.5 mean, 0.2 std.}
\psfrag{Cumulative probability}[c][c][0.8]{CDF $P(\log_{10}(\sigma_{\tau,h}[\text{s}]) < abscissa)$}
\psfrag{Data}[c][c][0.8]{$\log_{10}(\sigma_{\tau,h})[\text{s}]$ }
\subfigure[]{\includegraphics[width=0.47\textwidth]{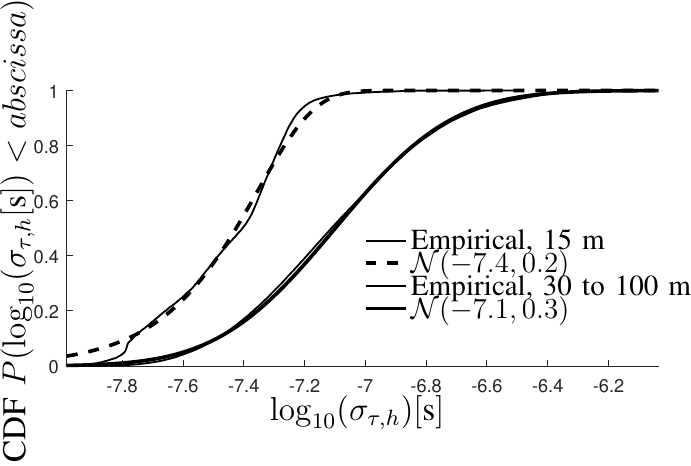}\label{horizontal_delay_spreads_cdfs}}
\caption{(a) RMS delay spreads denoted in logarithm vs. horizontal distance for the five horizontal flights. (b) CDFs of delay spreads denoted in logarithm for the five horizontal flights.}
\end{figure}

Fig.\ \ref{horizontal_delay_spreads} illustrates the delay spreads in logarithm scale vs. horizontal distance for the five horizontal flights. It can be observed that the characteristics of the delay spreads are similar to that of the K factors. For example, the delay spread at the height of 15 meters is obviously lower than that of the other four flights. {\blue We postulate that this is due to the downward BS antenna radiation pattern. Moreover, the delay spread at the height of 15 meters becomes lower with increasing horizontal distance.} This is because the channel far away from the BS is more LoS-alike with less MPCs as illustrated in Fig.\ \ref{sage_delay_75m}. {\blue However, for the heights above 15 meters, the delay spreads can be larger when the UAV is at the far end of the route due to the NLoS paths with larger delay and the weakness of the LoS path.} Fig.\ \ref{horizontal_delay_spreads_cdfs} illustrates the CDFs of the delay spreads in logarithm scale for the horizontal flights. ${\mathcal{N}(-7.4,0.2)}$ is found to roughly fit the empirical CDF at the height of 15 meters, and the CDF for the other four heights is best fitted by ${\mathcal{N}(-7.1,0.3)}$. In addition, the correlation coefficients between the logarithm delay spread and the horizontal distance are calculated as -0.76 and -0.38, respectively.

\begin{figure}
	\centering
\psfrag{data1}[l][l][0.7]{Height of 15 m}
\psfrag{data2}[l][l][0.7]{Height of 30 m}
\psfrag{data3}[l][l][0.7]{Height of 50 m}
\psfrag{data4}[l][l][0.7]{Height of 75 m}
\psfrag{data5}[l][l][0.7]{Height of 100 m}
\psfrag{dopplerspread}[c][c][0.8]{$\log_{10}(\sigma_{\nu,h})[\text{Hz}]$}
\psfrag{distance}[c][c][0.8]{Horizontal distance [m]}
\subfigure[]{\includegraphics[width=0.47\textwidth]{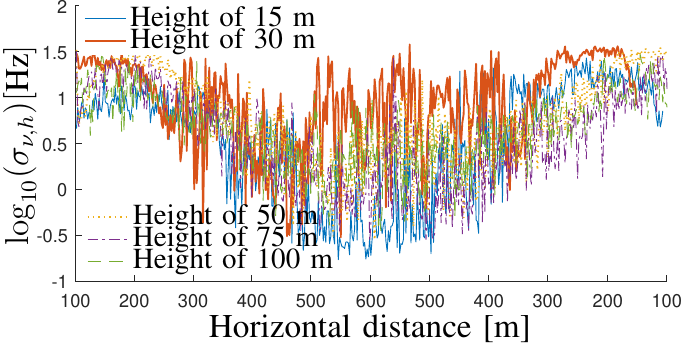}\label{horizontal_doppler_spreads}}
\psfrag{dopplerspreads15to100}[l][l][0.7]{Empirical, 15 to 100 m}
\psfrag{fit 9}[l][l][0.7]{ ${\text{EV}(0.9,0.4)}$}
\psfrag{Cumulative probability}[c][c][0.8]{CDF $P(\log_{10}(\sigma_{\nu,h}[\text{Hz}]) < abscissa)$}
\psfrag{Data}[c][c][0.8]{$\log_{10}(\sigma_{\nu,h})[\text{Hz}]$ }
\subfigure[]{\includegraphics[width=0.47\textwidth]{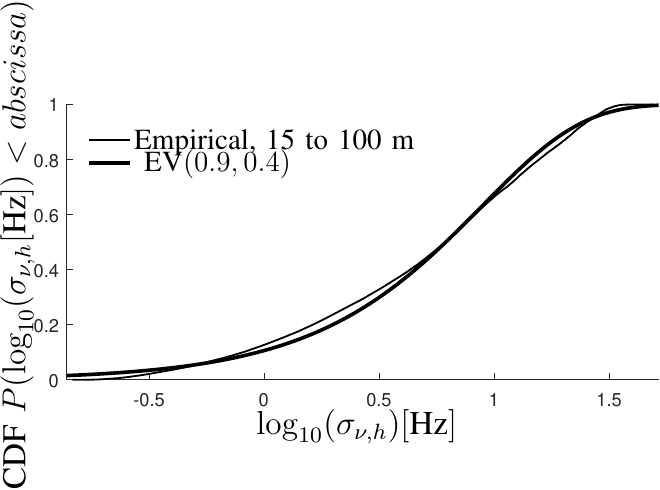}\label{horizontal_doppler_spreads_cdfs}}
\caption{(a) Logarithm Doppler spreads variation for the five horizontal flights. (b) CDF of Doppler frequency spreads denoted in logarithm for the five horizontal flights.}
\end{figure}

Fig.\ \ref{horizontal_doppler_spreads} illustrates the RMS Doppler frequency spreads for the five horizontal flights. Due to the fact that the speed of the UAV is not so large, the Doppler frequency spreads for the five horizontal flights are similar. The CDF for all the Doppler frequency spreads is illustrated in Fig.\ \ref{horizontal_doppler_spreads_cdfs} and is fitted with extreme value distribution\footnote{The probability density function of
$\text{EV}(\mu,\sigma)$ is formatted as $ f(x)= \sigma^{-1}\exp(\frac{x-\mu}{\sigma})\exp(-\exp(\frac{x-\mu}{\sigma}))$.} ${\text{EV}(0.9,0.4)}$. Moreover, the correlation coefficient with respect to horizontal distance is calculates as -0.55.

\subsection{Vertical flights}

\begin{figure}
	\centering
	\psfrag{data1}[l][l][0.7]{Position 1}
	\psfrag{data2}[l][l][0.7]{Position 2}
	\psfrag{data3}[l][l][0.7]{Position 3}
	\psfrag{data4}[l][l][0.7]{Position 4}
	\psfrag{data5}[l][l][0.7]{Position 5}
	\psfrag{verdis}[c][c][0.8]{Height [m]}
	\psfrag{power}[c][c][0.8]{Relative power [dB]}
	\includegraphics[width=0.45\textwidth]{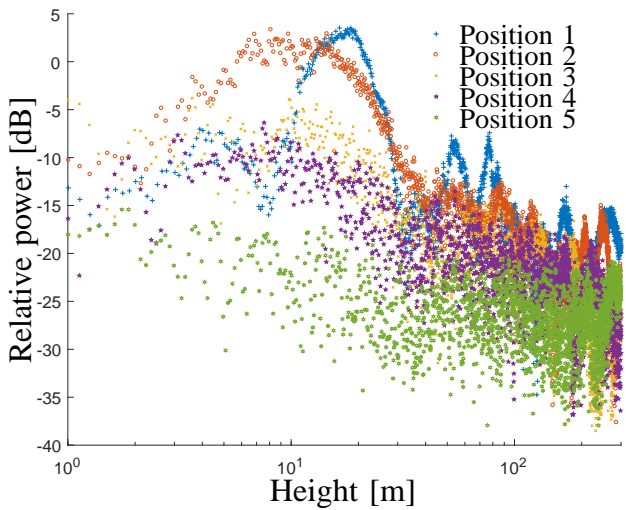}
	\caption{Received power for the five vertical flights at markers 1 to 5.\label{vertical_flights_power}}
\end{figure}

Fig.\ \ref{vertical_flights_power} illustrates the received power for the five vertical flights at the five ascending positions, respectively. We have the following preliminary findings: \textit{i)} In general trend, the received power decreases with the height increasing for one vertical flight. It is straightforward since the link distance gets larger; \textit{ii)} In general trend, the power decay slope with respect to the height becomes lower at a further ascending point; \textit{iii)} Fast fading becomes severer when the horizontal distance is larger.
The investigations and explanations for the characteristics are as follows.

\subsubsection{Path loss}
\begin{figure}
	\centering
    \psfrag{distance}[c][c][0.8]{Height [m]}
	\psfrag{power}[c][c][0.8]{Relative power [dB]}
	\psfrag{data1}[l][l][0.7]{Empirical, position 1}
    \psfrag{data2}[l][l][0.7]{Fitted}
    \psfrag{data3}[l][l][0.7]{Empirical, position 2}
    \psfrag{data4}[l][l][0.7]{Fitted}
   \psfrag{data5}[l][l][0.7]{Empirical, position 3}
   \psfrag{data6}[l][l][0.7]{Fitted}
     \psfrag{data7}[l][l][0.7]{Empirical, position 4}
    \psfrag{data8}[l][l][0.7]{Fitted}
   \psfrag{data9}[l][l][0.7]{Empirical, position 5}
   \psfrag{data10}[l][l][0.7]{Fitted}
	\includegraphics[width=0.45\textwidth]{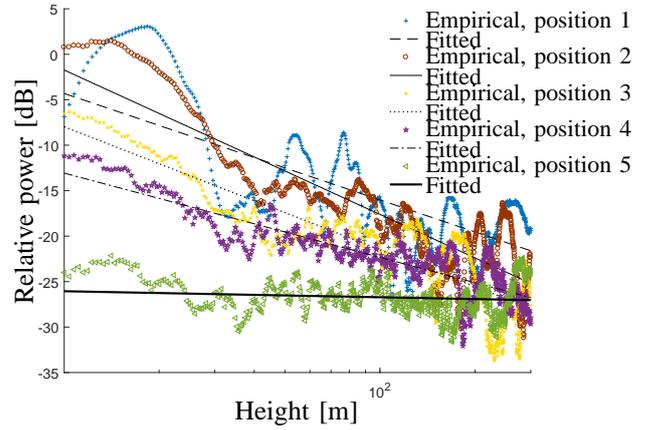}
	\caption{Power curve fitting for the five vertical flights. \label{path_loss_fitting_v}}
\end{figure}

Similarly to the case of horizontal flights, we propose a modified close-in free path loss model
%\begin{align}
%P_\mathrm{L}[dB]  = P_{\mathrm{L},d}(h_0) {+}10\gamma_{d} \cdot \log_{10}(\frac{h}{h_0})+ X_d \label{eq:vsplmodel}% + 10\gamma_{h} \cdot \log_{10}(\frac{d_h}{d_0})  \label{eq:splmodel}
%\end{align} where $h_0$ is the reference distance (1 m), $h$ is the height, $\gamma_{d}$ is the path loss exponent which is related to the horizontal distance $d$, and $X_d$ is the shadow fading.
{\red \begin{align}
P_\mathrm{L}[dB]  = 10\gamma_{d} \cdot \log_{10}(h)+ X_d + b_d \label{eq:vsplmodel}% + 10\gamma_{h} \cdot \log_{10}(\frac{d_h}{d_0})  \label{eq:splmodel}
\end{align} where $h$ is the height, $\gamma_{d}$ denotes the path loss exponent which is related to the horizontal distance $d$, and $X_d$ represents the shadow fading.}
Fig.\ \ref{path_loss_fitting_v} illustrates the smoothed power and the fittings for the five vertical flights. \addnew{It is noteworthy that the onboard antenna pattern could affect the results for the position 1 at large height values. In this case, the power would be reduced not only due to the distance, but also due to the Rx radiation pattern. For the other flights, the effect of the RX antenna pattern should be almost negligible.} Table \ref{tab:plev} presents the five path loss exponents obtained.
Mainly two factors have effects on the path loss exponent when the UAV flies up. One is that the link distance becomes larger, the other is that the UAV experiences different parts of the BS antenna radiation pattern. {\blue For example at position 5, the height increasing has little influence on the link distance and the crossing of BS antenna pattern, which results in a path loss exponent close to 0.}

%For example at position 5, the height has less effect both on the link distance as well as on the UAV elevation angle with respect to the BS, due to the horizontal distance is large. The power decay caused by the two factors is minor, which results in a path loss exponent close to 0.

\begin{table}
\centering
%\psfrag{Power angular spectra, [dB]}[l][l][0.6]{}
\caption{Path loss exponents for vertical flights}
\scalebox{0.9}{
\begin{tabular}{cccccc}
\hline
%\multicolumn{4}{c}{{{{\textit{Measurement specifications}}}}}\\\hline
Horizontal distance [m] &  100 & 200 & 300 & 400 & 500     \\
$\gamma_d$ &  1.17 & 1.58 & 1.35 & 0.92 & 0.07     \\ \hline
%Tx antenna height & $1.5$ meters & Rx antenna height & $1.9$ meters\\ \hline
\end{tabular}}
\label{tab:plev}
\end{table}

\subsubsection{Shadow fading}
\begin{figure}
	\centering
\psfrag{data1}[l][l][0.7]{Position 1}
\psfrag{data2}[l][l][0.7]{Position 2}
\psfrag{data3}[l][l][0.7]{Position 3}
\psfrag{data4}[l][l][0.7]{Position 4}
\psfrag{data5}[l][l][0.7]{Position 5}
\psfrag{shadow}[c][c][0.8]{$X_d$ [dB]}
\psfrag{distance}[l][l][0.7]{Height [m]}
\subfigure[]{\includegraphics[width=0.47\textwidth]{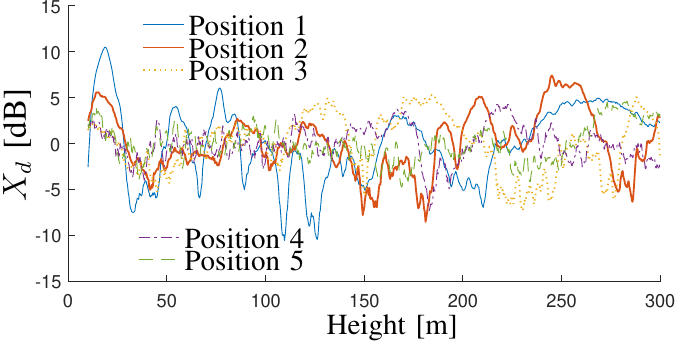}\label{v_shadowing_variation}}
\psfrag{shadow data}[l][l][0.7]{Empirical data, all positions}
\psfrag{3.0}[l][l][0.7]{${\mathcal{N}(0,3.0)}$}
\psfrag{Cumulative probability}[c][c][0.8]{CDF $P(X_d \text{[dB]} < abscissa)$}
\psfrag{Data}[c][c][0.8]{$X_d$ [dB]}
\subfigure[]{\includegraphics[width=0.47\textwidth]{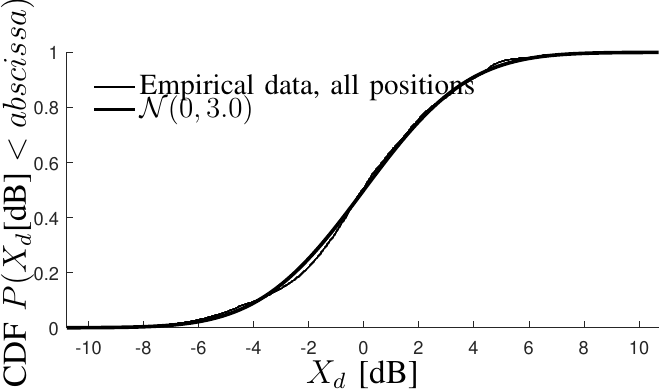}\label{v_shadow_fading_cdfs}}
\caption{(a) Shadow fading vs. height for the five vertical flights. (b) CDF of shadow fading for the five vertical flights. }
\end{figure}

Fig.\ \ref{v_shadowing_variation} illustrates the shadow fading for the five vertical flights. {\blue It can be observed that at positions 1 and 2, the shadow fading deviation at lower heights is larger. This is probably due to the blockage of UAV itself since the receiver antenna was fixed below the UAV.} Generally, the shadow fading for the five vertical flights is
%It can be observed that except for few cases at position 1, which we postulate is due to the blockage of the UAV itself, the shadow fading for the five vertical flights is
similar. For simplicity, we use the same statistical model for all of them. Fig.\ \ref{v_shadow_fading_cdfs} illustrates the empirical CDF and fitted CDF of ${\mathcal{N}(0,3.0)}$ for the shadow fading. Moreover, the correlation coefficient between the absolute shadow fading and the height is calculated to be 0.16.

\subsubsection{K factors}
\begin{figure}
	\centering
\psfrag{distance}[c][c][0.7]{Height [m]}
\psfrag{data1}[l][l][0.7]{Position 1}
\psfrag{data2}[l][l][0.7]{Position 2}
\psfrag{data3}[l][l][0.7]{Position 3}
\psfrag{data4}[l][l][0.7]{Position 4}
\psfrag{data5}[l][l][0.7]{Position 5}
\psfrag{kfactor}[c][c][0.8]{$K_d $ [dB]}
\subfigure[]{\includegraphics[width=0.47\textwidth]{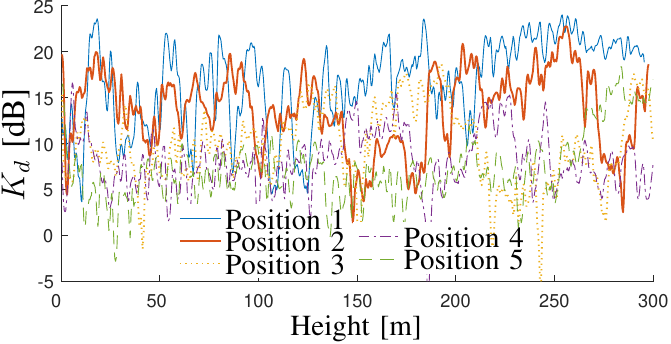}\label{v_fast_fading_variation}}
\psfrag{k12}[l][l][0.7]{Empirical, positions 1 and 2}
\psfrag{k345}[l][l][0.7]{Empirical, positions 3 to 5}
\psfrag{fit1}[l][l][0.7]{${\mathcal{N}(15.2,4.7)}$}
\psfrag{fit2}[l][l][0.7]{${\mathcal{N}(8.4,3.8)}$}
\psfrag{Cumulative probability}[c][c][0.8]{CDF $P(K_d[\text{dB}] < abscissa)$}
\psfrag{Data}[c][c][0.8]{$K_d$ [dB]}
\subfigure[]{\includegraphics[width=0.47\textwidth]{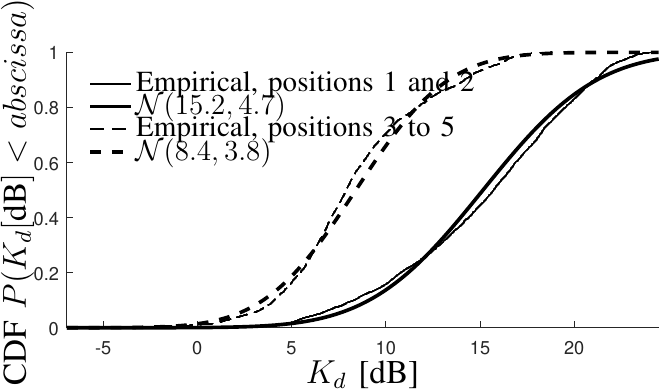}\label{v_fast_fading_cdfs}}
\caption{(a) K factor vs. height for the five vertical flights. (b) CDFs of K factors for the five vertical flights.}
\end{figure}

Fig.\ \ref{v_fast_fading_variation} illustrates the K factors for the five vertical flights. Observations and calculations show that the K factors for positions 1 and 2 are similar, while the K factors for positions 3 to 5 are also similar and lower than the previous ones. {\blue This is probably due to the fact when the ascending position is far away from the BS, the UAV is out of the main beam of BS for most of time. }% and is more likely to be blocked with increasing height.
 Fig.\ \ref{v_fast_fading_cdfs} illustrates the empirical CDFs for the K factors and analytical CDFs of ${\mathcal{N}(15.2,4.7)}$ and ${\mathcal{N}(8.4,3.8)}$ which are found to fit the empirical ones, respectively. In addition, the correlation coefficients with respect to the height are calculated as 0.29 and 0.20, which demonstrates that the K factor is likely to be larger with increasing  height. This is reasonable and intuitive as the channel above the sky is more LoS-alike.

\subsubsection{RMS delay spread and Doppler frequency spread}

\begin{figure}
	\centering
\psfrag{distance}[c][c][0.8]{Height [m]}
\psfrag{data1}[l][l][0.7]{Position 1}
\psfrag{data2}[l][l][0.7]{Position 2}
\psfrag{data3}[l][l][0.7]{Position 3}
\psfrag{data4}[l][l][0.7]{Position 4}
\psfrag{data5}[l][l][0.7]{Position 5}
\psfrag{delayspread}[c][c][0.8]{$\log_{10}(\sigma_{\tau,d})[\text{s}]$}
\subfigure[]{\includegraphics[width=0.47\textwidth]{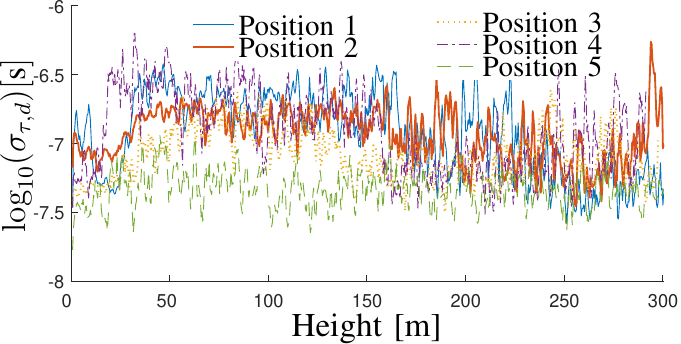}\label{vertical_delay_spreads}}
\psfrag{log_delay_spreads_5 data}[l][l][0.7]{Empirical, position 5}
\psfrag{delayspreads data}[l][l][0.7]{Empirical, positions 1 to 4}
\psfrag{-7.33,0.13}[l][l][0.7]{${\mathcal{N}(-7.33,0.13)}$}
\psfrag{-6.97,0.25}[l][l][0.7]{${\mathcal{N}(-6.97,0.25)}$}
\psfrag{Cumulative probability}[c][c][0.8]{CDF $P(\log_{10}(\sigma_{\tau,d}[\text{s}]) < abscissa)$}
\psfrag{Data}[c][c][0.8]{$\log_{10}(\sigma_{\tau,d}[\text{s}])$ }
\subfigure[]{\includegraphics[width=0.47\textwidth]{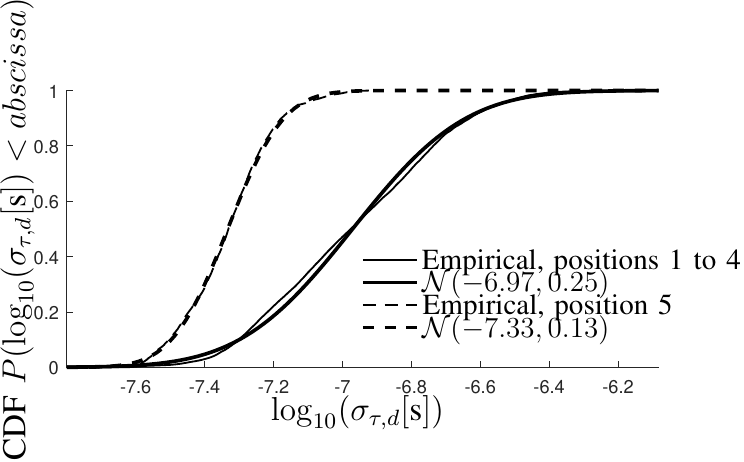}\label{vertical_delay_spreads_cdfs}}
\caption{(a) Delay spreads denoted in logarithm for the five vertical flights. (b) CDFs of delay spreads denoted in logarithm for the five vertical flights.}
\end{figure}

Fig.\ \ref{vertical_delay_spreads} illustrates the delay spreads in logarithm scale for the vertical flights. It can be observed that the delay spread for position 5 is obviously different from that for the other four positions. We postulate that it is mainly due to the scarcity of the scatterers, since the observed delay spread is also low in Fig.\ \ref{horizontal_delay_spreads} at the horizontal distance of 500 meters. For the other four positions, the delay spread values are larger. Moreover, basically the delay spreads for positions 1 to 4 decrease with the height increasing. %However, there are some cases of high delay spread for not low heights.
This is due to the fact the channel above the sky is more LoS-alike with less MPCs. {\blue However, the NLoS path is further from the LoS path as illustrated in Fig.\ \ref{sage_delay_vertical}, and the LoS path power can sometimes decay significantly since the UAV is out of the main beam, which can also result in large delay spreads in some high-height cases.} Fig.\ \ref{vertical_delay_spreads_cdfs} illustrates the empirical CDFs and fitted analytical CDFs of and for the delay spreads in logarithm scale at positions 1 to 4 and position 5, respectively. In addition, the correlation coefficients between the logarithm delay spread and the height are calculated as -0.38 and -0.12, respectively.

\begin{figure}
	\centering
\psfrag{distance}[c][c][0.8]{Height [m]}
\psfrag{data1}[l][l][0.7]{Position 1}
\psfrag{data2}[l][l][0.7]{Position 2}
\psfrag{data3}[l][l][0.7]{Position 3}
\psfrag{data4}[l][l][0.7]{Position 4}
\psfrag{data5}[l][l][0.7]{Position 5}
\psfrag{dopplerspread}[c][c][0.8]{$\log_{10}(\sigma_{\nu,d}[\text{Hz}])$}
\subfigure[]{\includegraphics[width=0.47\textwidth]{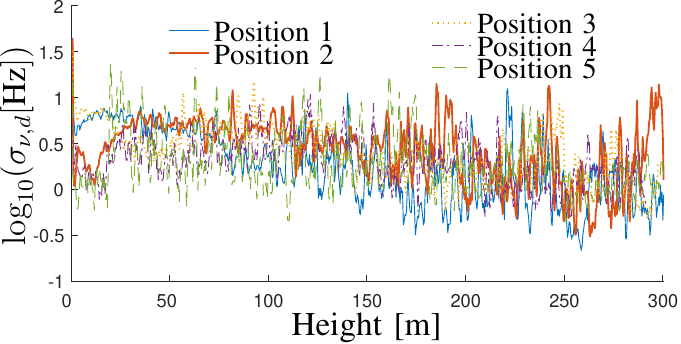}\label{vertical_doppler_spreads}}
\psfrag{CDF}[c][c][0.8]{CDF $P(\log_{10}(\sigma_{\nu,d}[\text{Hz}]) < abscissa)$}
\psfrag{Data}[c][c][0.8]{$\log_{10}(\sigma_{\nu,d}[\text{Hz}])$}
\psfrag{doppler_spreas data}[l][l][0.7]{Empirical, positions 1 to 5}
\psfrag{0.32,0.32}[l][l][0.7]{${\mathcal{N}(-0.32,0.32)}$}
\subfigure[]{\includegraphics[width=0.47\textwidth]{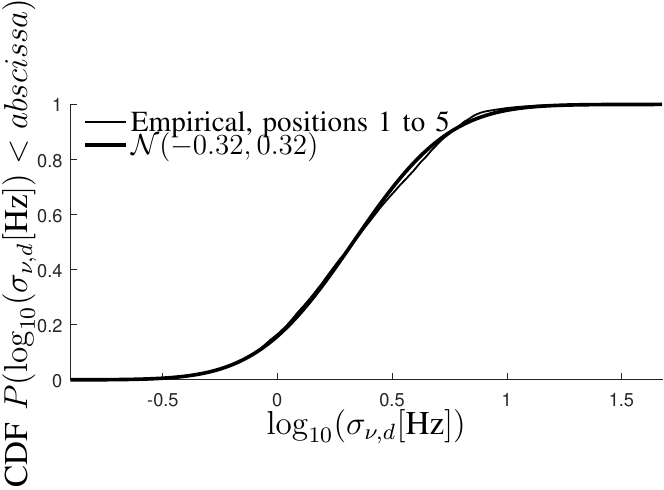}\label{v_doppler_spreads_cdfs}}
\caption{Doppler spreads denoted in logarithm for the five horizontal flights. (b) CDF of Doppler frequency spreads in logarithm for the five vertical flights.}
\end{figure}

Fig.\ \ref{vertical_doppler_spreads} illustrates the Doppler frequency spreads in logarithm scale for the five vertical flights. It can be observed that the Doppler frequency spreads are similar for different positions because the UAV speed is low. However, it is obvious that the Doppler frequency spreads become lower with increasing height. This is reasonable since the Doppler frequencies of all paths converge to the mimimum Doppler frequency with height increasing. For model simplicity, we use  a common statistical model for all of them. Fig.\ \ref{v_doppler_spreads_cdfs} illustrates the empirical CDF and the fitted analytical CDF of ${\mathcal{N}(-0.32,0.32)}$. In addition, the correlation coefficient between the logarithm Doppler frequency spread and the height is calculated as -0.7.

Table \ref{tab:horizontal_statistics} and Table \ref{tab:vertical_statistics} summarize the extracted statistics for the horizontal flights and vertical flights, respectively.

\begin{table}
	\caption{Statistics extracted for the five horizontal flights.}
	\begin{center}
\scriptsize $\rho_d$: correlation coefficient between parameter and horizontal distance
    \end{center}
        \hspace{0.1cm}
	\begin{center}
		\begin{tabular}{c|c|c|c|c|c}

            \hline
            \multirow{2}{*}{Parameter} & \multicolumn{5}{c}{(Mean, Std.); $\rho_d$ at different heights [m]} \\
            \cline{2-6}
              & 15  & 30  & 50  & 75  & 100    \\ \hline
            $\gamma_{h}$ & 3.64 & 2.30 & 2.28 & 1.31 & 1.67   \\ \hline
            $X_h$ [dB] & \multicolumn{5}{c}{(0, 2.7); 0.36}   \\ \hline
            $K_h$ [dB] & (12.6, 5.1); -0.64 & \multicolumn{4}{c}{(7.6, 5.6); -0.65}  \\ \hline
             $\log_{10}(\sigma_{\tau,h}[\text{s}])$ & (-7.41, 0.22); -0.76 & \multicolumn{4}{c}{(-7.12, 0.33); -0.38}  \\ \hline
            $\log_{10}(\sigma_{\nu,h}[\text{Hz}])$ &\multicolumn{5}{c}{(0.9, 0.4); -0.55}  \\ \hline
		\end{tabular}
	\end{center}
	\label{tab:horizontal_statistics}
\end{table}

\begin{table}
	\caption{Statistics extracted for the five vertical flights.}
	\begin{center}
\scriptsize $\rho_h$: correlation coefficient between parameter and height
    \end{center}

	\begin{center}
		\begin{tabular}{c|c|c|c|c|c}
			\hline
			\multirow{3}{*}{Parameter} & \multicolumn{5}{c}{(Mean, Std.); $\rho_h$  at different ascending positions} \\
			& \multicolumn{5}{c}{with different horizontal distances [m]} \\
			\cline{2-6}
			& 100  & 200  & 300  & 400  & 500   \\ \hline
			$\gamma_{d}$ & 1.17 & 1.58 & 1.35 & 0.92 & 0.07   \\ \hline
			$X_d$ [dB] & \multicolumn{5}{c}{(0, 3.0); 0.16}  \\ \hline
			$K_d $ [dB] & \multicolumn{2}{c|}{(15.2, 4.7); 0.29} & \multicolumn{3}{c}{(8.4, 3.8); 0.20}  \\ \hline
			$\log_{10}(\sigma_{\tau,d}[\text{s}])$ & \multicolumn{4}{c|}{(-6.97, 0.25); -0.38} & (-7.33, 0.13); -0.12     \\ \hline
			$\log_{10}(\sigma_{\nu,d}[\text{Hz}])$ &\multicolumn{5}{c}{(-0.3, 0.3); -0.7}   \\ \hline
		\end{tabular}
	\end{center}
	\label{tab:vertical_statistics}
\end{table}

\section{Conclusions} \label{conclusions}

In this contribution, the air-to-ground channel characteristics were investigated in a commercial long-time-evolution (LTE) network at the frequency band of 2.585 GHz. Five horizontal flights at different heights and five ascending flights with different horizontal distances to the base station (BS) were applied. The height of the BS was 20 meters. In the vertical flights, it is found that the channel is more line of sight (LoS) alike at a higher height with less multipath components (MPCs). Generally, K factors are positively correlated with the height, and delay spreads and Doppler frequency spreads are negatively correlated with the height. {\blue However, due to the downward radiation pattern of the LTE BS antenna, some opposite cases can be observed.} Furthermore, the horizontal distance of the ascending points also has impacts on the channel characteristics, e.g., the path loss exponent generally decreases with the horizontal distance increasing.

For the horizonal flights, it is found that the channel at the height of 15 meters is much different from the channels at the heights above, which is due to the variation of MPCs and the {\blue empirical radiation pattern of the LTE BS antenna.} Specifically, the path loss exponent is 3.64 at the height of 15 meters however decreases to less than 2 at the heights of 75 meters or 100 meters, the K factor at the height of 15 meters is 5 dB larger than that for the other above heights, and the delay spread at the height of 15 meters can be half of that for the other above heights. {\blue The channel statistics extracted in the model provide authentic and valuable reference to design and evaluate the future UAV applications integrated in the current LTE networks, as well as give insights on how to optimize an LTE network for supporting UAV-based applications.}

\bibliography{uav_channel_modeling_before_R1}
%\bibliography{uav_bib}
\end{document}